\documentclass[traditabstract,longauth]{aa} 
\usepackage{graphicx}
\usepackage{amsmath}
\usepackage{epsf}
\usepackage[breaklinks, colorlinks, citecolor=blue]{hyperref}
\usepackage{natbib}
\usepackage{amssymb}
\usepackage{color}
\usepackage{txfonts}
\bibpunct{(}{)}{;}{a}{}{,}

\def\setsymbol#1#2{\expandafter\def\csname #1\endcsname{#2}}
\def\getsymbol#1{\csname #1\endcsname}

\def\Planck{{\it Planck\/}}

\def\allearlypapers{\nocite{planck2011-1.1, planck2011-1.3, planck2011-1.4, planck2011-1.5, planck2011-1.6, planck2011-1.7, planck2011-1.10, planck2011-1.10sup, planck2011-5.1a, planck2011-5.1b, planck2011-5.2a, planck2011-5.2b, planck2011-5.2c, planck2011-6.1, planck2011-6.2, planck2011-6.3a, planck2011-6.4a, planck2011-6.4b, planck2011-6.6, planck2011-7.0, planck2011-7.2, planck2011-7.3, planck2011-7.7a, planck2011-7.7b, planck2011-7.12, planck2011-7.13}}

\newbox\tablebox    \newdimen\tablewidth
\def\leaderfil{\leaders\hbox to 5pt{\hss.\hss}\hfil}
\def\tablenote#1 #2\par{\begingroup \parindent=0.8em
    \abovedisplayshortskip=0pt\belowdisplayshortskip=0pt
    \noindent
    $$\hss\vbox{\hsize\tablewidth \hangindent=\parindent \hangafter=1 \noindent
    \hbox to \parindent{\sup{\rm #1}\hss}\strut#2\strut\par}\hss$$
    \endgroup}

\def\L2{\ifmmode L_2\else $L_2$\fi}

\def\DeltaT{\ifmmode \Delta T\else $\Delta T$\fi}
\def\deltat{\ifmmode \Delta t\else $\Delta t$\fi}
\def\fknee{\ifmmode f_{\rm knee}\else $f_{\rm knee}$\fi}
\def\Fmax{\ifmmode F_{\rm max}\else $F_{\rm max}$\fi}
\def\solar{\ifmmode{\rm M}_{\mathord\odot}\else${\rm M}_{\mathord\odot}$\fi}

\def\inv{\ifmmode^{-1}\else$^{-1}$\fi}
\def\mo{\ifmmode^{-1}\else$^{-1}$\fi}
\def\sup#1{\ifmmode ^{\rm #1}\else $^{\rm #1}$\fi}
\def\expo#1{\ifmmode \times 10^{#1}\else $\times 10^{#1}$\fi}
\def\,{\thinspace}
\def\lsim{\mathrel{\raise .4ex\hbox{\rlap{$<$}\lower 1.2ex\hbox{$\sim$}}}}
\def\gsim{\mathrel{\raise .4ex\hbox{\rlap{$>$}\lower 1.2ex\hbox{$\sim$}}}}

\def\simprop{\mathrel{\raise .4ex\hbox{\rlap{$\propto$}\lower 1.2ex\hbox{$\sim$}}}}
\def\deg{\ifmmode^\circ\else$^\circ$\fi}
\def\pdeg{\ifmmode $\setbox0=\hbox{$^{\circ}$}\rlap{\hskip.11\wd0 .}$^{\circ}
          \else \setbox0=\hbox{$^{\circ}$}\rlap{\hskip.11\wd0 .}$^{\circ}$\fi}
\def\arcs{\ifmmode {^{\scriptstyle\prime\prime}}
          \else $^{\scriptstyle\prime\prime}$\fi}
\def\arcm{\ifmmode {^{\scriptstyle\prime}}
          \else $^{\scriptstyle\prime}$\fi}
\newdimen\sa  \newdimen\sb
\def\parcs{\sa=.07em \sb=.03em
     \ifmmode \hbox{\rlap{.}}^{\scriptstyle\prime\kern -\sb\prime}\hbox{\kern -\sa}
     \else \rlap{.}$^{\scriptstyle\prime\kern -\sb\prime}$\kern -\sa\fi}
\def\parcm{\sa=.08em \sb=.03em
     \ifmmode \hbox{\rlap{.}\kern\sa}^{\scriptstyle\prime}\hbox{\kern-\sb}
     \else \rlap{.}\kern\sa$^{\scriptstyle\prime}$\kern-\sb\fi}
\def\ra[#1 #2 #3.#4]{#1\sup{h}#2\sup{m}#3\sup{s}\llap.#4}
\def\dec[#1 #2 #3.#4]{#1\deg#2\arcm#3\arcs\llap.#4}
\def\deco[#1 #2 #3]{#1\deg#2\arcm#3\arcs}
\def\rra[#1 #2]{#1\sup{h}#2\sup{m}}

\def\dots{\relax\ifmmode \ldots\else $\ldots$\fi}
\def\WHzsr{\ifmmode $W\,Hz\mo\,sr\mo$\else W\,Hz\mo\,sr\mo\fi}
\def\mHz{\ifmmode $\,mHz$\else \,mHz\fi}
\def\GHz{\ifmmode $\,GHz$\else \,GHz\fi}
\def\mKs{\ifmmode $\,mK\,s$^{1/2}\else \,mK\,s$^{1/2}$\fi}
\def\muKs{\ifmmode \,\mu$K\,s$^{1/2}\else \,$\mu$K\,s$^{1/2}$\fi}
\def\muKRJs{\ifmmode \,\mu$K$_{\rm RJ}$\,s$^{1/2}\else \,$\mu$K$_{\rm RJ}$\,s$^{1/2}$\fi}
\def\muKHz{\ifmmode \,\mu$K\,Hz$^{-1/2}\else \,$\mu$K\,Hz$^{-1/2}$\fi}
\def\MJysr{\ifmmode \,$MJy\,sr\mo$\else \,MJy\,sr\mo\fi}
\def\MJysrmK{\ifmmode \,$MJy\,sr\mo$\,mK$_{\rm CMB}\mo\else \,MJy\,sr\mo\,mK$_{\rm CMB}\mo$\fi}
\def\microns{\ifmmode \,\mu$m$\else \,$\mu$m\fi}

\def\muK{\ifmmode \,\mu$K$\else \,$\mu$\hbox{K}\fi}
\def\microK{\ifmmode \,\mu$K$\else \,$\mu$\hbox{K}\fi}
\def\muW{\ifmmode \,\mu$W$\else \,$\mu$\hbox{W}\fi}
\def\kms{\ifmmode $\,km\,s$^{-1}\else \,km\,s$^{-1}$\fi}
\def\kmsMpc{\ifmmode $\,\kms\,Mpc\mo$\else \,\kms\,Mpc\mo\fi}
\setsymbol{LFI:center:frequency:70GHz:units}{70.3\,GHz}
\setsymbol{LFI:center:frequency:44GHz:units}{44.1\,GHz}
\setsymbol{LFI:center:frequency:30GHz:units}{28.5\,GHz}

\setsymbol{LFI:center:frequency:70GHz}{70.3}
\setsymbol{LFI:center:frequency:44GHz}{44.1}
\setsymbol{LFI:center:frequency:30GHz}{28.5}

\setsymbol{LFI:center:frequency:LFI18:Rad:M:units}{71.7\GHz}
\setsymbol{LFI:center:frequency:LFI19:Rad:M:units}{67.5\GHz}
\setsymbol{LFI:center:frequency:LFI20:Rad:M:units}{69.2\GHz}
\setsymbol{LFI:center:frequency:LFI21:Rad:M:units}{70.4\GHz}
\setsymbol{LFI:center:frequency:LFI22:Rad:M:units}{71.5\GHz}
\setsymbol{LFI:center:frequency:LFI23:Rad:M:units}{70.8\GHz}
\setsymbol{LFI:center:frequency:LFI24:Rad:M:units}{44.4\GHz}
\setsymbol{LFI:center:frequency:LFI25:Rad:M:units}{44.0\GHz}
\setsymbol{LFI:center:frequency:LFI26:Rad:M:units}{43.9\GHz}
\setsymbol{LFI:center:frequency:LFI27:Rad:M:units}{28.3\GHz}
\setsymbol{LFI:center:frequency:LFI28:Rad:M:units}{28.8\GHz}
\setsymbol{LFI:center:frequency:LFI18:Rad:S:units}{70.1\GHz}
\setsymbol{LFI:center:frequency:LFI19:Rad:S:units}{69.6\GHz}
\setsymbol{LFI:center:frequency:LFI20:Rad:S:units}{69.5\GHz}
\setsymbol{LFI:center:frequency:LFI21:Rad:S:units}{69.5\GHz}
\setsymbol{LFI:center:frequency:LFI22:Rad:S:units}{72.8\GHz}
\setsymbol{LFI:center:frequency:LFI23:Rad:S:units}{71.3\GHz}
\setsymbol{LFI:center:frequency:LFI24:Rad:S:units}{44.1\GHz}
\setsymbol{LFI:center:frequency:LFI25:Rad:S:units}{44.1\GHz}
\setsymbol{LFI:center:frequency:LFI26:Rad:S:units}{44.1\GHz}
\setsymbol{LFI:center:frequency:LFI27:Rad:S:units}{28.5\GHz}
\setsymbol{LFI:center:frequency:LFI28:Rad:S:units}{28.2\GHz}

\setsymbol{LFI:center:frequency:LFI18:Rad:M}{71.7}
\setsymbol{LFI:center:frequency:LFI19:Rad:M}{67.5}
\setsymbol{LFI:center:frequency:LFI20:Rad:M}{69.2}
\setsymbol{LFI:center:frequency:LFI21:Rad:M}{70.4}
\setsymbol{LFI:center:frequency:LFI22:Rad:M}{71.5}
\setsymbol{LFI:center:frequency:LFI23:Rad:M}{70.8}
\setsymbol{LFI:center:frequency:LFI24:Rad:M}{44.4}
\setsymbol{LFI:center:frequency:LFI25:Rad:M}{44.0}
\setsymbol{LFI:center:frequency:LFI26:Rad:M}{43.9}
\setsymbol{LFI:center:frequency:LFI27:Rad:M}{28.3}
\setsymbol{LFI:center:frequency:LFI28:Rad:M}{28.8}
\setsymbol{LFI:center:frequency:LFI18:Rad:S}{70.1}
\setsymbol{LFI:center:frequency:LFI19:Rad:S}{69.6}
\setsymbol{LFI:center:frequency:LFI20:Rad:S}{69.5}
\setsymbol{LFI:center:frequency:LFI21:Rad:S}{69.5}
\setsymbol{LFI:center:frequency:LFI22:Rad:S}{72.8}
\setsymbol{LFI:center:frequency:LFI23:Rad:S}{71.3}
\setsymbol{LFI:center:frequency:LFI24:Rad:S}{44.1}
\setsymbol{LFI:center:frequency:LFI25:Rad:S}{44.1}
\setsymbol{LFI:center:frequency:LFI26:Rad:S}{44.1}
\setsymbol{LFI:center:frequency:LFI27:Rad:S}{28.5}
\setsymbol{LFI:center:frequency:LFI28:Rad:S}{28.2}

\setsymbol{LFI:white:noise:sensitivity:70GHz:units}{134.7\muKs}
\setsymbol{LFI:white:noise:sensitivity:44GHz:units}{164.7\muKs}
\setsymbol{LFI:white:noise:sensitivity:30GHz:units}{143.4\muKs}

\setsymbol{LFI:white:noise:sensitivity:70GHz}{134.7}
\setsymbol{LFI:white:noise:sensitivity:44GHz}{164.7}
\setsymbol{LFI:white:noise:sensitivity:30GHz}{143.4}

\setsymbol{LFI:white:noise:sensitivity:LFI18:Rad:M:units}{512.0\muKs}
\setsymbol{LFI:white:noise:sensitivity:LFI19:Rad:M:units}{581.4\muKs}
\setsymbol{LFI:white:noise:sensitivity:LFI20:Rad:M:units}{590.8\muKs}
\setsymbol{LFI:white:noise:sensitivity:LFI21:Rad:M:units}{455.2\muKs}
\setsymbol{LFI:white:noise:sensitivity:LFI22:Rad:M:units}{492.0\muKs}
\setsymbol{LFI:white:noise:sensitivity:LFI23:Rad:M:units}{507.7\muKs}
\setsymbol{LFI:white:noise:sensitivity:LFI24:Rad:M:units}{462.2\muKs}
\setsymbol{LFI:white:noise:sensitivity:LFI25:Rad:M:units}{413.6\muKs}
\setsymbol{LFI:white:noise:sensitivity:LFI26:Rad:M:units}{478.6\muKs}
\setsymbol{LFI:white:noise:sensitivity:LFI27:Rad:M:units}{277.7\muKs}
\setsymbol{LFI:white:noise:sensitivity:LFI28:Rad:M:units}{312.3\muKs}
\setsymbol{LFI:white:noise:sensitivity:LFI18:Rad:S:units}{465.7\muKs}
\setsymbol{LFI:white:noise:sensitivity:LFI19:Rad:S:units}{555.6\muKs}
\setsymbol{LFI:white:noise:sensitivity:LFI20:Rad:S:units}{623.2\muKs}
\setsymbol{LFI:white:noise:sensitivity:LFI21:Rad:S:units}{564.1\muKs}
\setsymbol{LFI:white:noise:sensitivity:LFI22:Rad:S:units}{534.4\muKs}
\setsymbol{LFI:white:noise:sensitivity:LFI23:Rad:S:units}{542.4\muKs}
\setsymbol{LFI:white:noise:sensitivity:LFI24:Rad:S:units}{399.2\muKs}
\setsymbol{LFI:white:noise:sensitivity:LFI25:Rad:S:units}{392.6\muKs}
\setsymbol{LFI:white:noise:sensitivity:LFI26:Rad:S:units}{418.6\muKs}
\setsymbol{LFI:white:noise:sensitivity:LFI27:Rad:S:units}{302.9\muKs}
\setsymbol{LFI:white:noise:sensitivity:LFI28:Rad:S:units}{285.3\muKs}

\setsymbol{LFI:white:noise:sensitivity:LFI18:Rad:M}{512.0}
\setsymbol{LFI:white:noise:sensitivity:LFI19:Rad:M}{581.4}
\setsymbol{LFI:white:noise:sensitivity:LFI20:Rad:M}{590.8}
\setsymbol{LFI:white:noise:sensitivity:LFI21:Rad:M}{455.2}
\setsymbol{LFI:white:noise:sensitivity:LFI22:Rad:M}{492.0}
\setsymbol{LFI:white:noise:sensitivity:LFI23:Rad:M}{507.7}
\setsymbol{LFI:white:noise:sensitivity:LFI24:Rad:M}{462.2}
\setsymbol{LFI:white:noise:sensitivity:LFI25:Rad:M}{413.6}
\setsymbol{LFI:white:noise:sensitivity:LFI26:Rad:M}{478.6}
\setsymbol{LFI:white:noise:sensitivity:LFI27:Rad:M}{277.7}
\setsymbol{LFI:white:noise:sensitivity:LFI28:Rad:M}{312.3}
\setsymbol{LFI:white:noise:sensitivity:LFI18:Rad:S}{465.7}
\setsymbol{LFI:white:noise:sensitivity:LFI19:Rad:S}{555.6}
\setsymbol{LFI:white:noise:sensitivity:LFI20:Rad:S}{623.2}
\setsymbol{LFI:white:noise:sensitivity:LFI21:Rad:S}{564.1}
\setsymbol{LFI:white:noise:sensitivity:LFI22:Rad:S}{534.4}
\setsymbol{LFI:white:noise:sensitivity:LFI23:Rad:S}{542.4}
\setsymbol{LFI:white:noise:sensitivity:LFI24:Rad:S}{399.2}
\setsymbol{LFI:white:noise:sensitivity:LFI25:Rad:S}{392.6}
\setsymbol{LFI:white:noise:sensitivity:LFI26:Rad:S}{418.6}
\setsymbol{LFI:white:noise:sensitivity:LFI27:Rad:S}{302.9}
\setsymbol{LFI:white:noise:sensitivity:LFI28:Rad:S}{285.3}

\setsymbol{LFI:knee:frequency:70GHz:units}{29.5\mHz}
\setsymbol{LFI:knee:frequency:44GHz:units}{56.2\mHz}
\setsymbol{LFI:knee:frequency:30GHz:units}{113.7\mHz}

\setsymbol{LFI:knee:frequency:70GHz}{29.5}
\setsymbol{LFI:knee:frequency:44GHz}{56.2}
\setsymbol{LFI:knee:frequency:30GHz}{113.7}

\setsymbol{LFI:knee:frequency:LFI18:Rad:M:units}{16.3\mHz}
\setsymbol{LFI:knee:frequency:LFI19:Rad:M:units}{15.1\mHz}
\setsymbol{LFI:knee:frequency:LFI20:Rad:M:units}{18.7\mHz}
\setsymbol{LFI:knee:frequency:LFI21:Rad:M:units}{37.2\mHz}
\setsymbol{LFI:knee:frequency:LFI22:Rad:M:units}{12.7\mHz}
\setsymbol{LFI:knee:frequency:LFI23:Rad:M:units}{34.6\mHz}
\setsymbol{LFI:knee:frequency:LFI24:Rad:M:units}{46.2\mHz}
\setsymbol{LFI:knee:frequency:LFI25:Rad:M:units}{24.9\mHz}
\setsymbol{LFI:knee:frequency:LFI26:Rad:M:units}{67.6\mHz}
\setsymbol{LFI:knee:frequency:LFI27:Rad:M:units}{187.4\mHz}
\setsymbol{LFI:knee:frequency:LFI28:Rad:M:units}{122.2\mHz}
\setsymbol{LFI:knee:frequency:LFI18:Rad:S:units}{17.7\mHz}
\setsymbol{LFI:knee:frequency:LFI19:Rad:S:units}{22.0\mHz}
\setsymbol{LFI:knee:frequency:LFI20:Rad:S:units}{8.7\mHz}
\setsymbol{LFI:knee:frequency:LFI21:Rad:S:units}{25.9\mHz}
\setsymbol{LFI:knee:frequency:LFI22:Rad:S:units}{15.8\mHz}
\setsymbol{LFI:knee:frequency:LFI23:Rad:S:units}{129.8\mHz}
\setsymbol{LFI:knee:frequency:LFI24:Rad:S:units}{100.9\mHz}
\setsymbol{LFI:knee:frequency:LFI25:Rad:S:units}{38.9\mHz}
\setsymbol{LFI:knee:frequency:LFI26:Rad:S:units}{58.9\mHz}
\setsymbol{LFI:knee:frequency:LFI27:Rad:S:units}{104.4\mHz}
\setsymbol{LFI:knee:frequency:LFI28:Rad:S:units}{40.7\mHz}

\setsymbol{LFI:knee:frequency:LFI18:Rad:M}{16.3}
\setsymbol{LFI:knee:frequency:LFI19:Rad:M}{15.1}
\setsymbol{LFI:knee:frequency:LFI20:Rad:M}{18.7}
\setsymbol{LFI:knee:frequency:LFI21:Rad:M}{37.2}
\setsymbol{LFI:knee:frequency:LFI22:Rad:M}{12.7}
\setsymbol{LFI:knee:frequency:LFI23:Rad:M}{34.6}
\setsymbol{LFI:knee:frequency:LFI24:Rad:M}{46.2}
\setsymbol{LFI:knee:frequency:LFI25:Rad:M}{24.9}
\setsymbol{LFI:knee:frequency:LFI26:Rad:M}{67.6}
\setsymbol{LFI:knee:frequency:LFI27:Rad:M}{187.4}
\setsymbol{LFI:knee:frequency:LFI28:Rad:M}{122.2}
\setsymbol{LFI:knee:frequency:LFI18:Rad:S}{17.7}
\setsymbol{LFI:knee:frequency:LFI19:Rad:S}{22.0}
\setsymbol{LFI:knee:frequency:LFI20:Rad:S}{8.7}
\setsymbol{LFI:knee:frequency:LFI21:Rad:S}{25.9}
\setsymbol{LFI:knee:frequency:LFI22:Rad:S}{15.8}
\setsymbol{LFI:knee:frequency:LFI23:Rad:S}{129.8}
\setsymbol{LFI:knee:frequency:LFI24:Rad:S}{100.9}
\setsymbol{LFI:knee:frequency:LFI25:Rad:S}{38.9}
\setsymbol{LFI:knee:frequency:LFI26:Rad:S}{58.9}
\setsymbol{LFI:knee:frequency:LFI27:Rad:S}{104.4}
\setsymbol{LFI:knee:frequency:LFI28:Rad:S}{40.7}

\setsymbol{LFI:slope:70GHz:units}{$-1.03$\mHz}
\setsymbol{LFI:slope:44GHz:units}{$-0.89$\mHz}
\setsymbol{LFI:slope:30GHz:units}{$-0.87$\mHz}

\setsymbol{LFI:slope:70GHz}{$-1.03$}
\setsymbol{LFI:slope:44GHz}{$-0.89$}
\setsymbol{LFI:slope:30GHz}{$-0.87$}

\setsymbol{LFI:slope:LFI18:Rad:M:units}{$-1.04$\mHz}
\setsymbol{LFI:slope:LFI19:Rad:M:units}{$-1.09$\mHz}
\setsymbol{LFI:slope:LFI20:Rad:M:units}{$-0.69$\mHz}
\setsymbol{LFI:slope:LFI21:Rad:M:units}{$-1.56$\mHz}
\setsymbol{LFI:slope:LFI22:Rad:M:units}{$-1.01$\mHz}
\setsymbol{LFI:slope:LFI23:Rad:M:units}{$-0.96$\mHz}
\setsymbol{LFI:slope:LFI24:Rad:M:units}{$-0.83$\mHz}
\setsymbol{LFI:slope:LFI25:Rad:M:units}{$-0.91$\mHz}
\setsymbol{LFI:slope:LFI26:Rad:M:units}{$-0.95$\mHz}
\setsymbol{LFI:slope:LFI27:Rad:M:units}{$-0.87$\mHz}
\setsymbol{LFI:slope:LFI28:Rad:M:units}{$-0.88$\mHz}
\setsymbol{LFI:slope:LFI18:Rad:S:units}{$-1.15$\mHz}
\setsymbol{LFI:slope:LFI19:Rad:S:units}{$-1.00$\mHz}
\setsymbol{LFI:slope:LFI20:Rad:S:units}{$-0.95$\mHz}
\setsymbol{LFI:slope:LFI21:Rad:S:units}{$-0.92$\mHz}
\setsymbol{LFI:slope:LFI22:Rad:S:units}{$-1.01$\mHz}
\setsymbol{LFI:slope:LFI23:Rad:S:units}{$-0.95$\mHz}
\setsymbol{LFI:slope:LFI24:Rad:S:units}{$-0.73$\mHz}
\setsymbol{LFI:slope:LFI25:Rad:S:units}{$-1.16$\mHz}
\setsymbol{LFI:slope:LFI26:Rad:S:units}{$-0.79$\mHz}
\setsymbol{LFI:slope:LFI27:Rad:S:units}{$-0.82$\mHz}
\setsymbol{LFI:slope:LFI28:Rad:S:units}{$-0.91$\mHz}

\setsymbol{LFI:slope:LFI18:Rad:M}{$-1.04$}
\setsymbol{LFI:slope:LFI19:Rad:M}{$-1.09$}
\setsymbol{LFI:slope:LFI20:Rad:M}{$-0.69$}
\setsymbol{LFI:slope:LFI21:Rad:M}{$-1.56$}
\setsymbol{LFI:slope:LFI22:Rad:M}{$-1.01$}
\setsymbol{LFI:slope:LFI23:Rad:M}{$-0.96$}
\setsymbol{LFI:slope:LFI24:Rad:M}{$-0.83$}
\setsymbol{LFI:slope:LFI25:Rad:M}{$-0.91$}
\setsymbol{LFI:slope:LFI26:Rad:M}{$-0.95$}
\setsymbol{LFI:slope:LFI27:Rad:M}{$-0.87$}
\setsymbol{LFI:slope:LFI28:Rad:M}{$-0.88$}
\setsymbol{LFI:slope:LFI18:Rad:S}{$-1.15$}
\setsymbol{LFI:slope:LFI19:Rad:S}{$-1.00$}
\setsymbol{LFI:slope:LFI20:Rad:S}{$-0.95$}
\setsymbol{LFI:slope:LFI21:Rad:S}{$-0.92$}
\setsymbol{LFI:slope:LFI22:Rad:S}{$-1.01$}
\setsymbol{LFI:slope:LFI23:Rad:S}{$-0.95$}
\setsymbol{LFI:slope:LFI24:Rad:S}{$-0.73$}
\setsymbol{LFI:slope:LFI25:Rad:S}{$-1.16$}
\setsymbol{LFI:slope:LFI26:Rad:S}{$-0.79$}
\setsymbol{LFI:slope:LFI27:Rad:S}{$-0.82$}
\setsymbol{LFI:slope:LFI28:Rad:S}{$-0.91$}

\setsymbol{LFI:FWHM:70GHz:units}{13\parcm01}
\setsymbol{LFI:FWHM:44GHz:units}{27\parcm92}
\setsymbol{LFI:FWHM:30GHz:units}{32\parcm65}

\setsymbol{LFI:FWHM:70GHz}{13.01}
\setsymbol{LFI:FWHM:44GHz}{27.92}
\setsymbol{LFI:FWHM:30GHz}{32.65}

\setsymbol{LFI:FWHM:LFI18:units}{13\parcm39}
\setsymbol{LFI:FWHM:LFI19:units}{13\parcm01}
\setsymbol{LFI:FWHM:LFI20:units}{12\parcm75}
\setsymbol{LFI:FWHM:LFI21:units}{12\parcm74}
\setsymbol{LFI:FWHM:LFI22:units}{12\parcm87}
\setsymbol{LFI:FWHM:LFI23:units}{13\parcm27}
\setsymbol{LFI:FWHM:LFI24:units}{22\parcm98}
\setsymbol{LFI:FWHM:LFI25:units}{30\parcm46}
\setsymbol{LFI:FWHM:LFI26:units}{30\parcm31}
\setsymbol{LFI:FWHM:LFI27:units}{32\parcm65}
\setsymbol{LFI:FWHM:LFI28:units}{32\parcm66}

\setsymbol{LFI:FWHM:LFI18}{13.39}
\setsymbol{LFI:FWHM:LFI19}{13.01}
\setsymbol{LFI:FWHM:LFI20}{12.75}
\setsymbol{LFI:FWHM:LFI21}{12.74}
\setsymbol{LFI:FWHM:LFI22}{12.87}
\setsymbol{LFI:FWHM:LFI23}{13.27}
\setsymbol{LFI:FWHM:LFI24}{22.98}
\setsymbol{LFI:FWHM:LFI25}{30.46}
\setsymbol{LFI:FWHM:LFI26}{30.31}
\setsymbol{LFI:FWHM:LFI27}{32.65}
\setsymbol{LFI:FWHM:LFI28}{32.66}

\setsymbol{LFI:FWHM:uncertainty:LFI18:units}{0.170\arcm}
\setsymbol{LFI:FWHM:uncertainty:LFI19:units}{0.174\arcm}
\setsymbol{LFI:FWHM:uncertainty:LFI20:units}{0.170\arcm}
\setsymbol{LFI:FWHM:uncertainty:LFI21:units}{0.156\arcm}
\setsymbol{LFI:FWHM:uncertainty:LFI22:units}{0.164\arcm}
\setsymbol{LFI:FWHM:uncertainty:LFI23:units}{0.171\arcm}
\setsymbol{LFI:FWHM:uncertainty:LFI24:units}{0.652\arcm}
\setsymbol{LFI:FWHM:uncertainty:LFI25:units}{1.075\arcm}
\setsymbol{LFI:FWHM:uncertainty:LFI26:units}{1.131\arcm}
\setsymbol{LFI:FWHM:uncertainty:LFI27:units}{1.266\arcm}
\setsymbol{LFI:FWHM:uncertainty:LFI28:units}{1.287\arcm}

\setsymbol{LFI:FWHM:uncertainty:LFI18}{0.170}
\setsymbol{LFI:FWHM:uncertainty:LFI19}{0.174}
\setsymbol{LFI:FWHM:uncertainty:LFI20}{0.170}
\setsymbol{LFI:FWHM:uncertainty:LFI21}{0.156}
\setsymbol{LFI:FWHM:uncertainty:LFI22}{0.164}
\setsymbol{LFI:FWHM:uncertainty:LFI23}{0.171}
\setsymbol{LFI:FWHM:uncertainty:LFI24}{0.652}
\setsymbol{LFI:FWHM:uncertainty:LFI25}{1.075}
\setsymbol{LFI:FWHM:uncertainty:LFI26}{1.131}
\setsymbol{LFI:FWHM:uncertainty:LFI27}{1.266}
\setsymbol{LFI:FWHM:uncertainty:LFI28}{1.287}

\setsymbol{HFI:center:frequency:100GHz:units}{100\,GHz}
\setsymbol{HFI:center:frequency:143GHz:units}{143\,GHz}
\setsymbol{HFI:center:frequency:217GHz:units}{217\,GHz}
\setsymbol{HFI:center:frequency:353GHz:units}{353\,GHz}
\setsymbol{HFI:center:frequency:545GHz:units}{545\,GHz}
\setsymbol{HFI:center:frequency:857GHz:units}{857\,GHz}

\setsymbol{HFI:center:frequency:100GHz}{100}
\setsymbol{HFI:center:frequency:143GHz}{143}
\setsymbol{HFI:center:frequency:217GHz}{217}
\setsymbol{HFI:center:frequency:353GHz}{353}
\setsymbol{HFI:center:frequency:545GHz}{545}
\setsymbol{HFI:center:frequency:857GHz}{857}

\setsymbol{HFI:Ndetectors:100GHz}{8}
\setsymbol{HFI:Ndetectors:143GHz}{11}
\setsymbol{HFI:Ndetectors:217GHz}{12}
\setsymbol{HFI:Ndetectors:353GHz}{12}
\setsymbol{HFI:Ndetectors:545GHz}{3}
\setsymbol{HFI:Ndetectors:857GHz}{4}

\setsymbol{HFI:FWHM:Maps:100GHz:units}{9\parcm88}
\setsymbol{HFI:FWHM:Maps:143GHz:units}{7\parcm18}
\setsymbol{HFI:FWHM:Maps:217GHz:units}{4\parcm87}
\setsymbol{HFI:FWHM:Maps:353GHz:units}{4\parcm65}
\setsymbol{HFI:FWHM:Maps:545GHz:units}{4\parcm72}
\setsymbol{HFI:FWHM:Maps:857GHz:units}{4\parcm39}
\setsymbol{HFI:FWHM:Maps:100GHz}{9.88}
\setsymbol{HFI:FWHM:Maps:143GHz}{7.18}
\setsymbol{HFI:FWHM:Maps:217GHz}{4.87}
\setsymbol{HFI:FWHM:Maps:353GHz}{4.65}
\setsymbol{HFI:FWHM:Maps:545GHz}{4.72}
\setsymbol{HFI:FWHM:Maps:857GHz}{4.39}

\setsymbol{HFI:beam:ellipticity:Maps:100GHz}{1.15}
\setsymbol{HFI:beam:ellipticity:Maps:143GHz}{1.01}
\setsymbol{HFI:beam:ellipticity:Maps:217GHz}{1.06}
\setsymbol{HFI:beam:ellipticity:Maps:353GHz}{1.05}
\setsymbol{HFI:beam:ellipticity:Maps:545GHz}{1.14}
\setsymbol{HFI:beam:ellipticity:Maps:857GHz}{1.19}

\setsymbol{HFI:FWHM:Mars:100GHz:units}{9\parcm37}
\setsymbol{HFI:FWHM:Mars:143GHz:units}{7\parcm04}
\setsymbol{HFI:FWHM:Mars:217GHz:units}{4\parcm68}
\setsymbol{HFI:FWHM:Mars:353GHz:units}{4\parcm43}
\setsymbol{HFI:FWHM:Mars:545GHz:units}{3\parcm80}
\setsymbol{HFI:FWHM:Mars:857GHz:units}{3\parcm67}

\setsymbol{HFI:FWHM:Mars:100GHz}{9.37}
\setsymbol{HFI:FWHM:Mars:143GHz}{7.04}
\setsymbol{HFI:FWHM:Mars:217GHz}{4.68}
\setsymbol{HFI:FWHM:Mars:353GHz}{4.43}
\setsymbol{HFI:FWHM:Mars:545GHz}{3.80}
\setsymbol{HFI:FWHM:Mars:857GHz}{3.67}

\setsymbol{HFI:beam:ellipticity:Mars:100GHz}{1.18}
\setsymbol{HFI:beam:ellipticity:Mars:143GHz}{1.03}
\setsymbol{HFI:beam:ellipticity:Mars:217GHz}{1.14}
\setsymbol{HFI:beam:ellipticity:Mars:353GHz}{1.09}
\setsymbol{HFI:beam:ellipticity:Mars:545GHz}{1.25}
\setsymbol{HFI:beam:ellipticity:Mars:857GHz}{1.03}

\setsymbol{HFI:CMB:relative:calibration:100GHz}{$\lsim 1\%$}
\setsymbol{HFI:CMB:relative:calibration:143GHz}{$\lsim 1\%$}
\setsymbol{HFI:CMB:relative:calibration:217GHz}{$\lsim 1\%$}
\setsymbol{HFI:CMB:relative:calibration:353GHz}{$\lsim 1\%$}
\setsymbol{HFI:CMB:relative:calibration:545GHz}{}
\setsymbol{HFI:CMB:relative:calibration:857GHz}{}

\setsymbol{HFI:CMB:absolute:calibration:100GHz}{$\lsim 2\%$}
\setsymbol{HFI:CMB:absolute:calibration:143GHz}{$\lsim 2\%$}
\setsymbol{HFI:CMB:absolute:calibration:217GHz}{$\lsim 2\%$}
\setsymbol{HFI:CMB:absolute:calibration:353GHz}{$\lsim 2\%$}
\setsymbol{HFI:CMB:absolute:calibration:545GHz}{}
\setsymbol{HFI:CMB:absolute:calibration:857GHz}{}

\setsymbol{HFI:FIRAS:gain:calibration:accuracy:statistical:100GHz}{}
\setsymbol{HFI:FIRAS:gain:calibration:accuracy:statistical:143GHz}{}
\setsymbol{HFI:FIRAS:gain:calibration:accuracy:statistical:217GHz}{}
\setsymbol{HFI:FIRAS:gain:calibration:accuracy:statistical:353GHz}{2.5\%}
\setsymbol{HFI:FIRAS:gain:calibration:accuracy:statistical:545GHz}{1\%}
\setsymbol{HFI:FIRAS:gain:calibration:accuracy:statistical:857GHz}{0.5\%}

\setsymbol{HFI:FIRAS:gain:calibration:accuracy:systematic:100GHz}{}
\setsymbol{HFI:FIRAS:gain:calibration:accuracy:systematic:143GHz}{}
\setsymbol{HFI:FIRAS:gain:calibration:accuracy:systematic:217GHz}{}
\setsymbol{HFI:FIRAS:gain:calibration:accuracy:systematic:353GHz}{}
\setsymbol{HFI:FIRAS:gain:calibration:accuracy:systematic:545GHz}{7\%}
\setsymbol{HFI:FIRAS:gain:calibration:accuracy:systematic:857GHz}{7\%}

\setsymbol{HFI:FIRAS:zero:point:accuracy:100GHz:units}{0.8\MJysr}
\setsymbol{HFI:FIRAS:zero:point:accuracy:143GHz:units}{}
\setsymbol{HFI:FIRAS:zero:point:accuracy:217GHz:units}{}
\setsymbol{HFI:FIRAS:zero:point:accuracy:353GHz:units}{1.4\MJysr}
\setsymbol{HFI:FIRAS:zero:point:accuracy:545GHz:units}{2.2\MJysr}
\setsymbol{HFI:FIRAS:zero:point:accuracy:857GHz:units}{1.7\MJysr}

\setsymbol{HFI:FIRAS:zero:point:accuracy:100GHz}{0.8}
\setsymbol{HFI:FIRAS:zero:point:accuracy:143GHz}{}
\setsymbol{HFI:FIRAS:zero:point:accuracy:217GHz}{}
\setsymbol{HFI:FIRAS:zero:point:accuracy:353GHz}{1.4}
\setsymbol{HFI:FIRAS:zero:point:accuracy:545GHz}{2.2}
\setsymbol{HFI:FIRAS:zero:point:accuracy:857GHz}{1.7}

\setsymbol{HFI:unit:conversion:100GHz:units}{0.2415\MJysrmK}
\setsymbol{HFI:unit:conversion:143GHz:units}{0.3694\MJysrmK}
\setsymbol{HFI:unit:conversion:217GHz:units}{0.4811\MJysrmK}
\setsymbol{HFI:unit:conversion:353GHz:units}{0.2883\MJysrmK}
\setsymbol{HFI:unit:conversion:545GHz:units}{0.05826\MJysrmK}
\setsymbol{HFI:unit:conversion:857GHz:units}{0.002238\MJysrmK}

\setsymbol{HFI:unit:conversion:100GHz}{0.2415}
\setsymbol{HFI:unit:conversion:143GHz}{0.3694}
\setsymbol{HFI:unit:conversion:217GHz}{0.4811}
\setsymbol{HFI:unit:conversion:353GHz}{0.2883}
\setsymbol{HFI:unit:conversion:545GHz}{0.05826}
\setsymbol{HFI:unit:conversion:857GHz}{0.002238}

\setsymbol{HFI:colour:correction:alpha=-2:V1.01:100GHz}{0.9893}
\setsymbol{HFI:colour:correction:alpha=-2:V1.01:143GHz}{0.9759}
\setsymbol{HFI:colour:correction:alpha=-2:V1.01:217GHz}{1.0007}
\setsymbol{HFI:colour:correction:alpha=-2:V1.01:353GHz}{1.0028}
\setsymbol{HFI:colour:correction:alpha=-2:V1.01:545GHz}{1.0019}
\setsymbol{HFI:colour:correction:alpha=-2:V1.01:857GHz}{0.9889}

\setsymbol{HFI:colour:correction:alpha=0:V1.01:100GHz}{1.0008}
\setsymbol{HFI:colour:correction:alpha=0:V1.01:143GHz}{1.0148}
\setsymbol{HFI:colour:correction:alpha=0:V1.01:217GHz}{0.9909}
\setsymbol{HFI:colour:correction:alpha=0:V1.01:353GHz}{0.9888}
\setsymbol{HFI:colour:correction:alpha=0:V1.01:545GHz}{0.9878}
\setsymbol{HFI:colour:correction:alpha=0:V1.01:857GHz}{1.0014}

\newfont{\gwpfont}{cmssq8 scaled 1000}
\newcommand{\rexcess}{{\gwpfont REXCESS}}
\def\Planck{{\it Planck\/}}
\def \planck {\hbox{\it Planck\/}}

\begin{document}
\author{\small
Planck Collaboration:
N.~Aghanim\inst{47}
\and
M.~Arnaud\inst{58}
\and
M.~Ashdown\inst{56, 4}
\and
J.~Aumont\inst{47}
\and
C.~Baccigalupi\inst{69}
\and
A.~Balbi\inst{30}
\and
A.~J.~Banday\inst{76, 7, 63}
\and
R.~B.~Barreiro\inst{53}
\and
M.~Bartelmann\inst{75, 63}
\and
J.~G.~Bartlett\inst{3, 54}
% ...
}
\institute{\small
Aalto University Mets\"{a}hovi Radio Observatory, Mets\"{a}hovintie 114, FIN-02540 Kylm\"{a}l\"{a}, Finland\\
\and
Agenzia Spaziale Italiana Science Data Center, c/o ESRIN, via Galileo Galilei, Frascati, Italy\\
\and
Astroparticule et Cosmologie, CNRS (UMR7164), Universit\'{e} Denis Diderot Paris 7, B\^{a}timent Condorcet, 10 rue A. Domon et L\'{e}onie Duquet, Paris, France\\
\and
Astrophysics Group, Cavendish Laboratory, University of Cambridge, J J Thomson Avenue, Cambridge CB3 0HE, U.K.\\
\and
Atacama Large Millimeter/submillimeter Array, ALMA Santiago Central Offices, Alonso de Cordova 3107, Vitacura, Casilla 763 0355, Santiago, Chile\\
\and
CITA, University of Toronto, 60 St. George St., Toronto, ON M5S 3H8, Canada\\
\and
CNRS, IRAP, 9 Av. colonel Roche, BP 44346, F-31028 Toulouse cedex 4, France\\
\and
California Institute of Technology, Pasadena, California, U.S.A.\\
\and
Centre of Mathematics for Applications, University of Oslo, Blindern, Oslo, Norway\\
\and
Centro de Astrof\'{\i}sica, Universidade do Porto, Rua das Estrelas, 4150-762 Porto, Portugal\\
\and
DAMTP, University of Cambridge, Centre for Mathematical Sciences, Wilberforce Road, Cambridge CB3 0WA, U.K.\\
\and
DSM/Irfu/SPP, CEA-Saclay, F-91191 Gif-sur-Yvette Cedex, France\\
\and
DTU Space, National Space Institute, Juliane Mariesvej 30, Copenhagen, Denmark\\
\and
Departamento de F\'{\i}sica, Universidad de Oviedo, Avda. Calvo Sotelo s/n, Oviedo, Spain\\
\and
Department of Astronomy and Astrophysics, University of Toronto, 50 Saint George Street, Toronto, Ontario, Canada\\
\and
Department of Physics \& Astronomy, University of British Columbia, 6224 Agricultural Road, Vancouver, British Columbia, Canada\\
\and
Department of Physics and Astronomy, University of Southern California, Los Angeles, California, U.S.A.\\
\and
Department of Physics, Gustaf H\"{a}llstr\"{o}min katu 2a, University of Helsinki, Helsinki, Finland\\
\and
Department of Physics, Princeton University, Princeton, New Jersey, U.S.A.\\
\and
Department of Physics, Purdue University, 525 Northwestern Avenue, West Lafayette, Indiana, U.S.A.\\
\and
Department of Physics, University of California, Berkeley, California, U.S.A.\\
\and
Department of Physics, University of California, One Shields Avenue, Davis, California, U.S.A.\\
\and
Department of Physics, University of California, Santa Barbara, California, U.S.A.\\
\and
Department of Physics, University of Illinois at Urbana-Champaign, 1110 West Green Street, Urbana, Illinois, U.S.A.\\
\and
Dipartimento di Fisica G. Galilei, Universit\`{a} degli Studi di Padova, via Marzolo 8, 35131 Padova, Italy\\
\and
Dipartimento di Fisica, Universit\`{a} La Sapienza, P. le A. Moro 2, Roma, Italy\\
\and
Dipartimento di Fisica, Universit\`{a} degli Studi di Milano, Via Celoria, 16, Milano, Italy\\
\and
Dipartimento di Fisica, Universit\`{a} degli Studi di Trieste, via A. Valerio 2, Trieste, Italy\\
\and
Dipartimento di Fisica, Universit\`{a} di Ferrara, Via Saragat 1, 44122 Ferrara, Italy\\
\and
Dipartimento di Fisica, Universit\`{a} di Roma Tor Vergata, Via della Ricerca Scientifica, 1, Roma, Italy\\
\and
Discovery Center, Niels Bohr Institute, Blegdamsvej 17, Copenhagen, Denmark\\
\and
Dpto. Astrof\'{i}sica, Universidad de La Laguna (ULL), E-38206 La Laguna, Tenerife, Spain\\
\and
European Southern Observatory, ESO Vitacura, Alonso de Cordova 3107, Vitacura, Casilla 19001, Santiago, Chile\\
\and
European Space Agency, ESAC, Planck Science Office, Camino bajo del Castillo, s/n, Urbanizaci\'{o}n Villafranca del Castillo, Villanueva de la Ca\~{n}ada, Madrid, Spain\\
\and
European Space Agency, ESTEC, Keplerlaan 1, 2201 AZ Noordwijk, The Netherlands\\
\and
Helsinki Institute of Physics, Gustaf H\"{a}llstr\"{o}min katu 2, University of Helsinki, Helsinki, Finland\\
\and
INAF - Osservatorio Astronomico di Padova, Vicolo dell'Osservatorio 5, Padova, Italy\\
\and
INAF - Osservatorio Astronomico di Roma, via di Frascati 33, Monte Porzio Catone, Italy\\
\and
INAF - Osservatorio Astronomico di Trieste, Via G.B. Tiepolo 11, Trieste, Italy\\
\and
INAF/IASF Bologna, Via Gobetti 101, Bologna, Italy\\
\and
INAF/IASF Milano, Via E. Bassini 15, Milano, Italy\\
\and
INRIA, Laboratoire de Recherche en Informatique, Universit\'{e} Paris-Sud 11, B\^{a}timent 490, 91405 Orsay Cedex, France\\
\and
IPAG: Institut de Plan\'{e}tologie et d'Astrophysique de Grenoble, Universit\'{e} Joseph Fourier, Grenoble 1 / CNRS-INSU, UMR 5274, Grenoble, F-38041, France\\
\and
Imperial College London, Astrophysics group, Blackett Laboratory, Prince Consort Road, London, SW7 2AZ, U.K.\\
\and
Infrared Processing and Analysis Center, California Institute of Technology, Pasadena, CA 91125, U.S.A.\\
\and
Institut N\'{e}el, CNRS, Universit\'{e} Joseph Fourier Grenoble I, 25 rue des Martyrs, Grenoble, France\\
\and
Institut d'Astrophysique Spatiale, CNRS (UMR8617) Universit\'{e} Paris-Sud 11, B\^{a}timent 121, Orsay, France\\
\and
Institut d'Astrophysique de Paris, CNRS UMR7095, Universit\'{e} Pierre \& Marie Curie, 98 bis boulevard Arago, Paris, France\\
\and
Institute of Astronomy and Astrophysics, Academia Sinica, Taipei, Taiwan\\
\and
Institute of Astronomy, University of Cambridge, Madingley Road, Cambridge CB3 0HA, U.K.\\
\and
Institute of Theoretical Astrophysics, University of Oslo, Blindern, Oslo, Norway\\
\and
Instituto de Astrof\'{\i}sica de Canarias, C/V\'{\i}a L\'{a}ctea s/n, La Laguna, Tenerife, Spain\\
\and
Instituto de F\'{\i}sica de Cantabria (CSIC-Universidad de Cantabria), Avda. de los Castros s/n, Santander, Spain\\
\and
Jet Propulsion Laboratory, California Institute of Technology, 4800 Oak Grove Drive, Pasadena, California, U.S.A.\\
\and
Jodrell Bank Centre for Astrophysics, Alan Turing Building, School of Physics and Astronomy, The University of Manchester, Oxford Road, Manchester, M13 9PL, U.K.\\
\and
Kavli Institute for Cosmology Cambridge, Madingley Road, Cambridge, CB3 0HA, U.K.\\
\and
LERMA, CNRS, Observatoire de Paris, 61 Avenue de l'Observatoire, Paris, France\\
\and
Laboratoire AIM, IRFU/Service d'Astrophysique - CEA/DSM - CNRS - Universit\'{e} Paris Diderot, B\^{a}t. 709, CEA-Saclay, F-91191 Gif-sur-Yvette Cedex, France\\
\and
Laboratoire Traitement et Communication de l'Information, CNRS (UMR 5141) and T\'{e}l\'{e}com ParisTech, 46 rue Barrault F-75634 Paris Cedex 13, France\\
\and
Laboratoire de Physique Subatomique et de Cosmologie, CNRS/IN2P3, Universit\'{e} Joseph Fourier Grenoble I, Institut National Polytechnique de Grenoble, 53 rue des Martyrs, 38026 Grenoble cedex, France\\
\and
Laboratoire de l'Acc\'{e}l\'{e}rateur Lin\'{e}aire, Universit\'{e} Paris-Sud 11, CNRS/IN2P3, Orsay, France\\
\and
Lawrence Berkeley National Laboratory, Berkeley, California, U.S.A.\\
\and
Max-Planck-Institut f\"{u}r Astrophysik, Karl-Schwarzschild-Str. 1, 85741 Garching, Germany\\
\and
Max-Planck-Institut f\"{u}r Extraterrestrische Physik, Giessenbachstra{\ss}e, 85748 Garching, Germany\\
\and
MilliLab, VTT Technical Research Centre of Finland, Tietotie 3, Espoo, Finland\\
\and
National University of Ireland, Department of Experimental Physics, Maynooth, Co. Kildare, Ireland\\
\and
Niels Bohr Institute, Blegdamsvej 17, Copenhagen, Denmark\\
\and
Observational Cosmology, Mail Stop 367-17, California Institute of Technology, Pasadena, CA, 91125, U.S.A.\\
\and
SISSA, Astrophysics Sector, via Bonomea 265, 34136, Trieste, Italy\\
\and
SUPA, Institute for Astronomy, University of Edinburgh, Royal Observatory, Blackford Hill, Edinburgh EH9 3HJ, U.K.\\
\and
School of Physics and Astronomy, Cardiff University, Queens Buildings, The Parade, Cardiff, CF24 3AA, U.K.\\
\and
Space Research Institute (IKI), Russian Academy of Sciences, Profsoyuznaya Str, 84/32, Moscow, 117997, Russia\\
\and
Space Sciences Laboratory, University of California, Berkeley, California, U.S.A.\\
\and
Stanford University, Dept of Physics, Varian Physics Bldg, 382 Via Pueblo Mall, Stanford, California, U.S.A.\\
\and
Universit\"{a}t Heidelberg, Institut f\"{u}r Theoretische Astrophysik, Albert-\"{U}berle-Str. 2, 69120, Heidelberg, Germany\\
\and
Universit\'{e} de Toulouse, UPS-OMP, IRAP, F-31028 Toulouse cedex 4, France\\
% ...
}

\title{\textit{Planck} early results. X. Statistical analysis of Sunyaev-Zeldovich scaling relations for X-ray galaxy clusters\thanks{Corresponding author: R. Piffaretti, \url{rocco.piffaretti@cea.fr}}}
\abstract{
All-sky data from the \planck \ survey and the Meta-Catalogue of X-ray detected Clusters of galaxies (MCXC) are combined to investigate the relationship between the thermal Sunyaev-Zeldovich (SZ) signal and X-ray luminosity. The sample comprises $\sim$ 1600 X-ray clusters with redshifts up to $\sim$ 1 and spans a wide range in X-ray luminosity. The SZ signal is extracted for each object individually, and the statistical significance of the measurement is maximised by averaging the SZ signal in bins of X-ray luminosity, total mass, or redshift. The SZ signal is detected at very high significance over more than two decades in X-ray luminosity ($10^{43} {\rm erg/s} \lesssim L_{\rm 500} E(z)^{-7/3} \lesssim 2 \times 10^{45} {\rm erg/s}$). The relation between intrinsic SZ signal and X-ray luminosity is investigated and the measured SZ signal is compared to values predicted from X-ray data. \planck \ measurements and X-ray based predictions are found to be in excellent agreement over the whole explored luminosity range. No significant deviation from standard evolution of the scaling relations is detected. For the first time the intrinsic scatter in the scaling relation between SZ signal and X-ray luminosity is measured and found to be consistent with the one in the luminosity -- mass relation from X-ray studies. There is no evidence of any deficit in SZ signal strength in \planck \ data relative to expectations from the X-ray properties of clusters, underlining the robustness and consistency of our overall view of intra-cluster medium properties.}
\keywords{Galaxy Clusters -- Large Scale Structure -- Planck}
\authorrunning{Planck Collaboration}
\titlerunning{Planck early results X.}
\maketitle
\allearlypapers
%
\section{Introduction}
\label{introduction.sec}

Clusters of galaxies are filled with a hot, ionised, intra-cluster medium (ICM) visible both in the X-ray band via thermal bremsstrahlung 
and from its distortion of the cosmic microwave background (CMB) from inverse Compton scattering, i.e., 
the Sunyaev-ZelÕdovich effect \citep[][SZ, hereafter]{SZ1970,SZ1972}. The SZ signal can be divided into a kinetic SZ and a thermal SZ effect, originating from bulk and thermal motions of ICM electrons, respectively. Since the kinetic SZ is a second-order effect, 
we only consider the thermal SZ effect. Because of the different scaling of SZ and X-ray fluxes with electron density and temperature, SZ and X-ray observations are highly complementary. The combination of information from these two types of observations is a powerful one for cosmological studies, as well as for improving our understanding of cluster physics \citep[see][for a review]{birkinshaw1999}. In this framework, it is paramount to investigate to what degree the ICM properties inferred from SZ and X-ray data agree. 

Unfortunately, there is currently no consensus on whether predictions for the SZ signal based on ICM properties
derived from X-ray observation agree with direct SZ observations, hampering our understanding of the involved physics. \cite{lieu2006} find evidence of a weaker SZ signal in the 3-year {\it WMAP\/} data than expected from {\it ROSAT\/} observations for 31 X-ray clusters. \cite{bielby2007} reach similar conclusions using the same {\it WMAP\/} data and ROSAT sample and the additional {\it Chandra\/} data for 38 clusters. Conversely, \cite{afshordi2007} find good agreement between the strength of the SZ signal in {\it WMAP\/} 3-year data and the X-ray properties of their sample of 193 massive galaxy clusters. The last findings are supported further by the results of \cite{atrio2008}, whose analysis is based on the same SZ data and a larger sample of 661 clusters. \cite{diego2010} argue that a large contamination from point sources is needed to reconcile the SZ signal seen
in the {\it WMAP\/} 5-year data with what is inferred from a large sample of {\it ROSAT\/} clusters. However, using the same SZ data and a slightly larger but similar sample of ROSAT clusters, \cite{melin2010} find good agreement between SZ signal and expectations. The latter finding is confirmed by the work by \cite{andersson2010}, where high quality {\it Chandra\/} data for 15 South Pole Telescope clusters are used. Finally, the {\it WMAP\/} 7-year data analysis by \cite{komatsu2010} argues for a deficit of SZ signal compared to expectations, especially at low masses.

Improved understanding of this issue is clearly desired, since it would provide invaluable knowledge about clusters of galaxies and aid in the interpretation and exploitation of SZ surveys such as the South Pole Telescope \citep[SPT,][]{carlstrom2009} survey, the Atacama Cosmology Telescope \citep[ACT,][]{fowler2007}, and \planck\footnote{\Planck\ (http://www.esa.int/\Planck ) is a project of the European Space Agency (ESA) with instruments provided by two scientific consortia funded by ESA member states (in particular the lead countries France and Italy), with contributions from NASA (USA) and telescope reflectors provided by a collaboration between ESA and a scientific consortium led and funded by Denmark.} 
\citep{tauber2010a}. Since August 2009 the \planck\ satellite has been surveying the whole sky in nine frequency bands with high sensitivity and a relatively high spatial resolution. \planck\ data thus offer the unique opportunity to fully explore this heavily debated issue.  As part of a series of papers on \planck\ early results on clusters of galaxies \citep{planck2011-5.1a,planck2011-5.1b,planck2011-5.2a,planck2011-5.2b,planck2011-5.2c}, we present a study of the relationship between X-ray luminosity and SZ signal in the direction of $\sim$ 1600 objects from the MCXC X-ray clusters compilation \citep{MCXC} and demonstrate that there is excellent agreement between SZ signal and expectations from the X-ray properties of clusters.

The paper is organised as follows. In Sect. \ref{data.sec} we briefly describe the \planck \ data used in the analysis and present the adopted X-ray sample. In Sect. \ref{clustermodel.sec} we present the baseline model used in the paper. The model description is rather comprehensive because the model is also adopted in the companion papers on \Planck\ early results on clusters of galaxies \citep{planck2011-5.1a,planck2011-5.1b,planck2011-5.2a,planck2011-5.2c}. Sect. \ref{SZextraction.sec} describes how the SZ signal is extracted from \planck \ frequency maps at the position of each MCXC cluster and how these are averaged in X-ray luminosity bins. Our results are presented in Sect. \ref{scalingrelations.sec} and robustness tests are detailed in Sect. \ref{robustness.sec}. Our findings are discussed and summarised in \ref{conclusions.sec}.

When necessary we adopt a $\Lambda$CDM cosmology with $H_{0} = 70$~ km/s/Mpc, $\Omega_{\rm M} = 0.3$ and $\Omega_\Lambda= 0.7$ throughout the paper. The quantity $E(z)$ is the ratio of the Hubble constant at redshift $z$ to its present value, $H_{0}$, i.e., $E(z)^2=\Omega_\mathrm{m} (1+z)^3 + \Omega_\mathrm{\Lambda}$. 

The total cluster mass $M_{\rm 500}$ is defined as the mass within  the radius $R_{\rm 500}$ within which the mean mass density is 500 times the critical density of the universe, $ \rho_{\rm crit}(z)$, at the cluster redshift: $M_{500} = {4 \over 3} \pi  \, \rho_{\rm crit}(z) \, 500 \, R_{500}^3$.  We adopt an overdensity of 500 since $R_{\rm 500}$ encloses a substantial fraction of the total virialised mass of the system while being the largest radius probed in current X-ray observations of large samples of galaxy clusters.

The SZ signal is characterised by $Y_{500}$ defined as $D^2_{\rm A}(z) \, Y_{500} = (\sigma_{\rm T}/m_{\rm e} c^2)\int P dV$, where $D_{\rm A}(z)$ is the angular distance to a system at redshift $z$,  $\sigma_{\rm T}$ is the Thomson cross-section, $c$ the speed of light,  $m_{\rm e}$ the electron rest mass, $P= n_{\rm e} k T_{\rm e}$ the pressure, defined as the product of the electron number density and temperature and the integration is performed over the sphere of radius $R_{500}$. The quantity $Y_{500}$ is proportional to the apparent magnitude of the SZ signal and  $D_{\rm A}^2 \, Y_{500}$ is the spherically integrated Compton parameter, which, for simplicity, will be referred to as SZ signal or {\it intrinsic\/} SZ signal in the remainder of the paper. All quoted X-ray luminosities are cluster rest frame luminosities, converted to the [0.1-2.4] keV band.

\section{Data}
\label{data.sec}

In the following subsections we present the \planck \ data and the X-ray cluster sample used in our analysis. In order to avoid contamination from galactic sources in the \planck \ data we exclude the galactic plane: $\mid b \mid \le 14 \, {\rm deg}$ from the maps.  In addition, we exclude clusters located less than $1.5 \times$ beam full width half maximum (FWHM) from point sources detected at more than $10\,\sigma$ in any of the single frequency \planck\ maps, because such sources can strongly affect SZ measurements. 
 
\subsection{SZ data set}
\label{SZdata.sec}

\Planck\ \citep{tauber2010a, planck2011-1.1} is the third generation space mission to measure the anisotropy of the CMB.  It observes the sky in nine frequency bands covering 30--857 \, GHz with high sensitivity and angular resolution from 31\arcm\ to 5\arcm.  The Low Frequency Instrument LFI; \citep{Mandolesi2010, Bersanelli2010, planck2011-1.4} covers the 30, 44, and 70 \, GHz bands with amplifiers cooled to 20\,\hbox{K}.  The High Frequency Instrument (HFI; \citealt{Lamarre2010, planck2011-1.5}) covers the 100, 143, 217, 353, 545, and 857\,GHz bands with bolometers cooled to 0.1\,\hbox{K}.  Polarisation is measured in all but the highest two bands \citep{Leahy2010, Rosset2010}.  A combination of radiative cooling and three mechanical coolers produces the temperatures needed for the detectors and optics \citep{planck2011-1.3}.  Two Data Processing Centers (DPCs) check and calibrate the data and make maps of the sky \citep{planck2011-1.7, planck2011-1.6}.  \Planck's sensitivity, angular resolution, and frequency coverage make it a powerful instrument for galactic and extragalactic astrophysics as well as cosmology. Early astrophysics results are given in Planck Collaboration, 2011e--x.

In this paper, we use only the six temperature channel maps of HFI (100, 143, 217, 353, 545 and 857 GHz), corresponding to (slightly more than) the first sky survey of Planck. Details of how these maps are produced can be found in \cite{planck2011-1.7, planck2011-5.1a}. At this early stage of the Planck SZ analysis adding the LFI channel maps does not bring significant improvements to our results. We use the full resolution maps at HEALPix\footnote{http://healpix.jpl.nasa.gov} nside=2048 (pixel size 1.72\arcm) and we assume that beams are adequately described by symmetric Gaussians with FWHM as given in Table \ref{tab:fwhm}. Uncertainties in our results due to beam corrections, map calibrations and uncertainties in bandpasses are small,
as shown in Sect. \ref{robustness.sec} below.
     \begin{table}
      \caption[]{{\footnotesize Values of the beam full width half maximum assumed for each of the six channel maps of HFI.}}
         \label{tab:fwhm}
        \begin{center}
        \begin{tabular}{l|cccccc}
         \hline
           \hline
         frequency [GHz] &  100 & 143 & 217& 353 & 545 & 857  \\
            \hline
          FWHM [\arcm]  & 9.53  & 7.08 &  4.71 & 4.50 & 4.72 & 4.42 \\
           \hline
           FWHM error [\arcm] & 0.10 & 0.12 &  0.17 & 0.14 & 0.21 & 0.28 \\
             \hline
        \end{tabular}
        \end{center}
   \end{table}
\subsection{X-ray data set}
\label{X-raydata.sec}

The cluster sample adopted in our analysis, the MCXC (Meta-Catalogue of X-ray detected Clusters of galaxies), is presented in detail in \cite{MCXC}. The information provided by all publicly available ROSAT All Sky Survey-based (NORAS: \cite{bohringer2000}, REFLEX: \cite{bohringer2004}, BCS: \cite{ebeling1998,ebeling2000}, SGP: \cite{cruddace2002}, NEP: \cite{henry2006}, MACS: \cite{ebeling2007,ebeling2010}, and CIZA: \cite{ebeling2002,kocevski2007}) and serendipitous (160SD: \cite{mullis2003}, 400SD: \cite{burenin2007}, SHARC: \cite{romer2000,burke2003}, WARPS: \cite{perlman2002,horner2008}, and EMSS: \cite{emss1994,henry2004}) cluster catalogues was systematically homogenised and duplicate entries were carefully handled, yielding a large catalogue of approximately $1800$ clusters. For each cluster the MCXC provides, among other quantities, coordinates, redshifts, and luminosities. The latter are central to the MCXC and to our analysis because luminosity is the only available mass proxy for such a large number of X-ray clusters. For this reason we will focus here on how the cluster rest frame luminosities provided by the MCXC are computed. Other quantities such as total mass and cluster size will be discussed in Sect. \ref{clustermodel.sec} below, because they are more model dependent. 

In addition to being converted to the cosmology adopted in this paper and to the [0.1-2.4] keV band (the typical X-ray survey energy band), luminosities are converted to that for an overdensity of 500 (see below). This allows us to minimise the scatter originating from the fact that publicly available catalogues provide luminosity measurements within different apertures. 

Because cluster catalogues generally provide luminosities measured within some aperture or luminosities extrapolated up to large radii (total luminosities), the luminosities $L_{\rm 500}$ provided by the MCXC were computed by converting the total luminosities to $L_{\rm 500}$ using a constant factor or, when aperture luminosities are available, by performing an iterative computation based on the \rexcess \ mean gas density profile and  $L_{\rm 500}$ -- $M_{\rm 500}$ relation. The \rexcess \  $L_{\rm 500}$ -- $M_{\rm 500}$ calibration is discussed in Sect. \ref{clustermodel.sec} below. While the comparison presented in Section 5.3 of \cite{MCXC} indicates that the differences between these two methods  do not introduce any systematic bias, it is clear that the iteratively computed  $L_{\rm 500}$ are the most accurate. The iterative computation was possible for the NORAS/REFLEX, BCS, SHARC, and NEP catalogues. As shown in \cite{MCXC} the luminosities $L_{\rm 500}$ depend very weakly on the assumed $L_{\rm 500}$ -- $M_{\rm 500}$ relation. Nevertheless, when exploring the different $L_{\rm 500}$ -- $M_{\rm 500}$ relations detailed below, we consistently recompute $L_{\rm 500}$ using the relevant $L_{\rm 500}$ -- $M_{\rm 500}$ relation. 

In addition, we supplement the MCXC sample with $z \ge 0.6$ cluster data in order to enlarge the redshift leverage. These additional high redshift clusters are collected from the literature by utilizing the X-Rays Clusters Database BAX \footnote{http://bax.ast.obs-mip.fr/} and performing a thorough search in the literature. For these objects we collect coordinates, redshift, and X-ray luminosity. The luminosity values given in the literature are converted to the [0.1-2.4] keV band and adopted cosmology as done in \cite{MCXC}. Because the available luminosities are derived under fairly different assumptions (e.g., aperture radius, extrapolation methods, etc.) we do not attempt to homogenise them to the fiducial luminosity $L_{\rm 500}$ as done in \cite{MCXC}. In almost all the cases the adopted luminosity is however either the total luminosity (i.e. extrapolated to large radii) or the directly the luminosity $L_{\rm 500}$. Given the fact that the difference between these is close to $10 \, \%$ and that uncertainties affecting luminosity measurement of high redshift clusters are much larger, we treat all luminosities as fiducial luminosity $L_{\rm 500}$. 

The MCXC and $z \ge 0.6$ supplementary clusters located around the galactic plane ($\mid b \mid \le 14 \, {\rm deg}$) or near bright point sources ($>10 \,  \sigma$, ${\rm distance} <1.5 \times {\rm FWHM}$) are excluded from the analysis.  The resulting sample comprises 1603 clusters, with 845 clusters being members of the NORAS/REFLEX sample. There is a total of 33 supplementary $z \ge 0.6$ clusters located in the sky region selected in our analysis. 

\begin{figure}
  \includegraphics[width=1. \columnwidth]{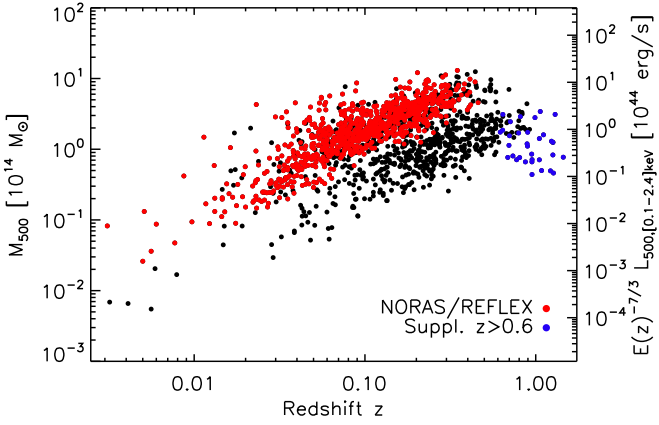}
  \caption{{\footnotesize Observed [0.1-2.4] keV band luminosities (right vertical axis) and inferred masses (left vertical axis) as a function of redshift. Shown are the MCXC (the NORAS/REFLEX control subsample in shown in red) and the supplementary clusters (blue dots).}}
  \label{sample:fig}
\end{figure}

In Fig. \ref{sample:fig} we show luminosity and mass as a function of redshift for the whole sample with the NORAS/REFLEX and  supplementary $z \ge 0.6$ clusters displayed with different colours. The figure shows the different clustering in the L-z plane of RASS (mostly NORAS/REFLEX) and serendipitously discovered clusters. The NORAS/REFLEX clusters are central to our study for many reasons. First, being the most luminous and numerous, they are expected to yield the bulk of the SZ signal from known clusters. Second, their distribution in the sky is uniform: NORAS and REFLEX cover the northern and southern sky, respectively, with the galactic plane excluded ($\mid b \mid \le 20 \, {\rm deg}$). Finally, the NORAS/REFLEX sample was also used in \cite{melin2010} in an analysis equivalent to the one presented in this work but based on {\it WMAP\/}-5yr data. For these reasons we use NORAS/REFLEX clusters as control sample in our analysis. 

For simplicity, in the remainder of the paper the whole compilation of MCXC plus supplementary $z \ge 0.6$ clusters will be referred to as MCXC. The [0.1-2.4] keV luminosities of the clusters in our sample range from $1.53 \times 10^{40}$ to $2.91\times 10^{45}$ erg s$^{-1}$, with a median luminosity of $0.95\times 10^{44}$ erg s$^{-1}$, and redshifts range from 0.0031 to 1.45. Notice that while the adopted sample essentially comprises all known X-ray clusters in the sky region of interest, its selection function is unknown. The latter issue and how we evaluate its impact on our results is discussed in Sect. \ref{LM.sec} below.
\section{The cluster model}
\label{clustermodel.sec}

Our cluster model is based on the \rexcess, a sample expressly designed to measure the structural and 
scaling properties of the local X-ray cluster population by means of an unbiased, representative sampling in 
luminosity \citep{bohringer2007}. The calibration of scaling relations and the average structural parameters of 
such an X-ray selected sample is ideal because is not morphologically biased. Furthermore, the gas properties of the \rexcess \ clusters 
can be traced by {\it XMM-Newton\/} up to large cluster-centric distances, allowing robust measurements at an 
overdensity of 500. 

Since X-ray luminosity is the only available mass proxy for our large cluster sample, the most fundamental 
ingredient of the cluster model is the scaling relation between [0.1-2.4] keV band luminosity and total cluster mass, 
which is detailed in Sect. \ref{LM.sec}. Given a cluster redshift $z$, mass $M_{\rm 500}$ and hence cluster size $R_{\rm 500}$, 
the universal pressure profile of  \cite{arnaud2010} is then used to predict the electronic pressure profile. This allows us to predict $D^2_{\rm A} \, Y_{\rm 500}$,  the SZ signal integrated in a sphere of radius $R_{\rm 500}$ as summarised in Sect. \ref{szsignal.sec}. It is important to notice that the estimated cluster size $R_{\rm 500}$ and the universal pressure profile are also assumed when extracting the SZ signal from \planck \ data as detailed in Sect. \ref{SZextraction.sec} below. 
 
In the following we describe the assumptions at the basis of our {\it fiducial model\/} and provide the adopted scaling laws. 
In addition, we also discuss how these assumptions are varied in order to investigate the robustness of our results. 

\subsection{$L_{\rm 500}$ -- $M_{\rm 500}$ relation}
\label{LM.sec}

For a given [0.1-2.4] keV band luminosity $L_{\rm 500}$ the total mass $M_{\rm 500}$ is estimated adopting the \rexcess\ $L_{\rm 500}$ -- $M_{\rm 500}$ relation \citep{pratt2009}:
\begin{equation}
E(z)^{-7/3} \, \left (  \frac{L_{500}}{10^{44} \, {\rm erg} \, {\rm s}^{-1}} \right ) =C_{LM} \, \left ( \frac{M_{500}}{3 \times 10^{14} \, M_{\odot}} \right )^{\alpha_{LM}} \, .
\label{L -- M:eq} 
\end{equation}
Because this relation has been calibrated using the low scatter X-ray mass proxy $Y_{\rm X}$ \citep{kravtsov2006}, the parameters $C_{\rm LM}$ and $\alpha_{\rm LM}$ depend on whether the slope of the underlying $M_{\rm 500}-Y_{\rm X}$ relation is assumed to be equal to the standard (self-similar) value of $\alpha_{\rm MY_{\rm X}}=3/5$ or it is allowed to be a free parameter, yielding $\alpha_{\rm MY_{\rm X}}=0.561$ \citep[see Eqs. 2 and 3 in][]{arnaud2010}. In the reminder of the paper these two cases will be referred to as {\it standard\/} and {\it empirical\/}, respectively. 
 Our {\it fiducial model\/} adopts the {\it empirical\/} case, which reflects the observed mass dependence of the gas mass fraction in galaxy clusters. It is thus fully observationally motivated.

In addition to these two variations of the $L_{\rm 500}$ -- $M_{\rm 500}$ relation, we also consider the impact of Malmquist bias on our analysis. To this end we perform our analysis using the L -- M calibrations derived from \rexcess\  luminosity data corrected or uncorrected for the Malmquist bias. In the reminder of the paper these two cases will be referred to as the {\it intrinsic\/} and \rexcess\  $L_{\rm 500}$ -- $M_{\rm 500}$ relations, respectively. Notice that the difference between the {\it intrinsic\/} and \rexcess\  $L_{\rm 500}$ -- $M_{\rm 500}$ relations is very small at high luminosity \citep{pratt2009}. Ideally, one should use the {\it intrinsic\/} $L_{\rm 500}$ -- $M_{\rm 500}$ relation and compute, for each sample used to construct the MCXC compilation, the observed $L_{\rm 500}$ -- $M_{\rm 500}$ relation according to each survey selection function. Unfortunately this would be possible only for a small fraction of MCXC clusters because the individual selection functions of the samples used to construct it are extremely complex and, in most of the cases, not known or not available. Therefore we simply consider the intrinsic $L_{\rm 500}$ -- $M_{\rm 500}$ relation as an extreme and illustrative case, since it is equivalent to assuming that selection effects of our X-ray sample are totally negligible. On the other hand, in particular for the NORAS/REFLEX control sample and at high luminosities, the \rexcess\  $L_{\rm 500}$ -- $M_{\rm 500}$ relation is expected to be quite close to the one that would be observed in our sample.  For these reasons, our {\it fiducial model\/} adopts the \rexcess\  $L_{\rm 500}$ -- $M_{\rm 500}$ relation and the {\it intrinsic\/} case is used to test the robustness of our results. 

These different choices result in four different calibrations of the $L_{\rm 500}$ -- $M_{\rm 500}$ relation. The corresponding best fitting parameters are summarised in Table \ref{tab:lx_m_param} \citep[see also][]{arnaud2010}. Values are given for the {\it fiducial case\/} where the observed \rexcess\  $L_{\rm X}-Y_{\rm X}$ and $M-Y_{\rm X}$ are assumed as well as for the cases where these two assumptions are varied: i.e. intrinsic (Malmquist bias corrected) $L_{\rm X}-Y_{\rm X}$ relation and standard slope of the $M-Y_{\rm X}$ relation $\alpha_{\rm MY_{\rm X}}$. The table also lists the intrinsic dispersion in each relation, which we use to investigate the effect of scatter in the assumed mass-observable relation in our analysis.

For a given $L_{\rm 500}$ -- $M_{\rm 500}$ relation we estimate, for each cluster in our sample, the total mass $M_{\rm 500}$ from its luminosity $L_{\rm 500}$. When the latter is computed iteratively (see Sect. \ref{X-raydata.sec}), the same $L_{\rm 500}$ -- $M_{\rm 500}$ relation is adopted for consistency. Finally, the cluster size or characteristic radius $R_{\rm 500}$ is computed from its definition: $M_{500} = {4 \over 3} \pi  \, \rho_{\rm crit}(z) \, 500 \, R_{500}^3$. 

     \begin{table}
      \caption[]{{\footnotesize Values for the parameters of the adopted $L_{\rm X}-M$ relation.}}
         \label{tab:lx_m_param}
        \begin{center}
        \begin{tabular}{ccccc}
         \hline
         \hline
          $\alpha_{\rm MY_{\rm X}}$& L -- M & $\log C_{LM} $ & $\alpha_{LM}$ & $\sigma_{\log L - \log M}$  \\
              \hline
              0.561 & \rexcess &  0.274 & 1.64 & 0.183 \\
                \hline
                 \hline
              0.561 &  Intrinsic & 0.193 & 1.76 & 0.199 \\
              3/5 & \rexcess  &  0.295 & 1.50 & 0.183 \\
              3/5 & Intrinsic & 0.215 & 1.61 & 0.199 \\
          \hline
        \end{tabular}
        \end{center}
   \end{table}

\subsection{The SZ signal}
\label{szsignal.sec}
As shown in \cite{arnaud2010}, if standard evolution is assumed, the average physical pressure profile of clusters can be described by 
 \begin{equation}
\label{cluster_profile}
     P(r) = P_{500} \left ( \frac{M_{500}}{3 \times 10^{14}\,M_{\odot}}\right ) ^{ \alpha_{\rm P}} {P_0 \over (c_{500} x)^\gamma (1+(c_{500} x)^\alpha)^{\frac{\beta-\gamma}{\alpha}}} \, ,
\end{equation}
with $x=r/ R_{\rm 500}$ and $\alpha_{\rm  P} = 1/ \alpha_{\rm MY_{\rm X}} - 5/3$. In the standard case we have $\alpha_{\rm  P} =0$, while in the empirical case $\alpha_{\rm  P} =0.12$. Notice that the most precise empirical description also takes into account a weak radial dependence of the exponent $\alpha_{\rm P}$ of the form $\alpha_{\rm P}=0.12+\alpha^{\prime}_{\rm P}(x)$. Here we neglect the radially dependent term since, as shown by \cite{arnaud2010}, it introduces a fully negligible correction. 

The characteristic pressure $P_{\rm 500}$ is defined as
\begin{equation}
\label{p500}
     P_{500} = 1.65 \times 10^{-3} E(z)^{8/3} \left ( {M_{500} \over 3\times10^{14}\, M_\odot} \right )^{2/3}   \; {\rm keV \, cm^{-3}}.
      \end{equation}
The set of parameters $[P_{0} ,c_{500},\gamma,\alpha,\beta]$ in Eq. \ref{cluster_profile} are constrained by fitting the \rexcess \ data and depend on the assumed slope of the $M-Y_{\rm X}$ relation. In Table \ref{tab:pressure_param} we list the adopted best fitting values, which, as detailed in Sect. \ref{SZextraction.sec} below, are also used to optimise the SZ signal detection. Values are first given for the {\it fiducial case} where the observed $M-Y_{\rm X}$ relation (with slope $\alpha_{\rm MY_{\rm X}}$=0.561) and the average profile of all \rexcess\ clusters are adopted. The values for the average cool-core (CC) and morphologically disturbed (MD)  \rexcess \ profiles, that we use to estimate the uncertainties originating from deviations from the average profile (see Sect. \ref{robustness.sec}), and the average profile derived assuming a standard slope of the $M-Y_{\rm X}$ relation ($\alpha_{\rm MY_{\rm X}}=3/5$), are also listed in the table.

Because of the large number of free parameters, there is a strong parameter degeneracy and therefore a comparison of individual parameters in Table \ref{tab:pressure_param} is meaningless. The parameters for the {\it standard\/} case are also listed in the table.
     \begin{table}
      \caption[]{{\footnotesize Parameters  describing the shape of the pressure profile.}}
         \label{tab:pressure_param}
        \begin{center}
        \begin{tabular}{ccccccc}
         \hline
         \hline
         & $\alpha_{\rm MY_{\rm X}}$ & $P_{0}$ & $c_{500}$ & $\gamma$ & $\alpha$ & $\beta$  \\
            \hline
          All  & 0.561 & 8.403 &  1.177 & 0.3081 & 1.0510 & 5.4905 \\
          \hline
           CC &  0.561 & 3.249 &  1.128  & 0.7736 & 1.2223 & 5.4905 \\
           MD & 0.561  & 3.202 &  1.083  & 0.3798 & 1.4063 & 5.4905 \\
                       \hline
                        \hline
          All  & 3/5  & 8.130 &  1.156 & 0.3292 & 1.0620 & 5.4807 \\
          \hline
        \end{tabular}
        \end{center}
   \end{table}

The model allows us to compute the physical pressure profile as a function of mass $M_{500}$ and $z$ and thus to obtain the $D^2_{\rm A} \, Y_{\rm 500}$ -- $M_{\rm 500}$ relation by integration of  $P(r)$ in Eq. \ref{cluster_profile} within a sphere of radius $R_{\rm 500}$. The relation can be written as
\begin{eqnarray}
\label{YM}
      D^2_{\rm A}(z) \, Y_{500} & = &  2.925\times10^{-5} I(1) \nonumber \\
  & \times &     \left ({M_{500} \over 3\times10^{14} \,M_\odot} \right )^{\frac{1}{ \alpha_{MY_X}}}   \; E(z)^{2/3}  \; {\rm Mpc}^2 
\end{eqnarray}
or, equivalently,
\begin{eqnarray}
\label{YM_arc}
Y_{500} & = &  1.383 \times10^{-3} I(1) \nonumber \\
  & \times & \left ({M_{500} \over 3\times10^{14} \,M_\odot} \right )^{\frac{1}{ \alpha_{MY_X}}}  E(z)^{2/3}   \;  \left ( {D_{\rm A}(z) \over 500 \, {\rm Mpc}} \right )^{-2} {\rm arcmin}^2 ,
\end{eqnarray}
where $I(1)=0.6145$ and $I(1)=0.6552$ are numerical factors arising from volume integrals of the pressure profile in the empirical and standard slope case, respectively \citep[see][for details]{arnaud2010}. Combining Eqs. \ref{L -- M:eq} and \ref{YM} gives
\begin{eqnarray}
\label{YL}
D^2_{\rm A}(z) \, Y_{500}  & = & 2.925\times10^{-5} I(1) \nonumber \\
& \times & \left [ \frac{E(z)^{-7/3}}{C_{LM}} \left ( \frac{L_{500}}{10^{44} \, {\rm erg} \, {\rm s} ^{-1}} \right) \right ]^{\frac{1}{\alpha_{LY}}} E(z)^{2/3}  {\rm Mpc}^2 ,
\end{eqnarray}
where $\alpha_{\rm LY}= \alpha_{\rm LM} \times \alpha_{\rm MY_X}$. In the {\it fiducial case\/} $\alpha_{\rm LY}= 0.92$, implying that $Y_{\rm 500} \, D^2_{\rm A} \propto L_{\rm 500}^{1.09}$ for our model predictions.
The cluster model allows us to predict the volume integrated Compton parameter $D^2_{\rm A} \, Y_{\rm 500}$ for each individual cluster in our large 
X-ray cluster sample from its [0.1-2.4] keV band luminosity $L_{\rm 500}$. These X-ray based prediction can be computed for different assumptions about the underlying X-ray scaling relations ({\it standard\/}/{\it empirical\/} and {\it intrinsic\/}/\rexcess\ cases) and compared with the observed SZ signal, whose measurement is detailed in the next section.       

To reiterate, our {\it fiducial case\/} assumes: {\it empirical\/} slope of $M-Y_{\rm X}$ relation and  \rexcess\ $L_{\rm 500}$ -- $M_{\rm 500}$ relation. If not otherwise stated, in the remainder of the paper results for the {\it fiducial case\/} are presented and results obtained by varying the assumptions are going to be compared to it in Sect. \ref{robustness.sec}.

For simplicity the cluster size and SZ signal for the MCXC clusters in \cite{planck2011-5.1a} are provided in the {\it standard\/} $M_{\rm 500}-Y_{\rm X}$ slope case.  As we show in \cite{planck2011-5.1a} the effects of this on X-ray size and both predicted and observed SZ quantities for clusters in the Early Sunyaev-Zeldovich (ESZ) catalog are fully negligible with respect to the overall uncertainties.

\section{Extraction of the SZ signal}
\label{SZextraction.sec}

\subsection{Individual measurements}

The SZ signal is extracted for each cluster individually by cutting from each of the six HFI frequency maps $10^\circ \times 10^\circ$ patches (pixel=1.72 arcmin) centred at the cluster position. The resulting set of six HFI frequency patches is then used to extract the cluster signal by means of multifrequency matched filters (MMF, hereafter). The main features of the multifrequency matched filters are summarised in \cite{melin2010} and more details can be found in \cite{herranz2002} and \cite{melin2006}.

The MMF algorithm optimally filters and combines the patches to estimate the SZ signal. It relies on an estimate of the noise auto- and cross-power-spectra from the patches. Working with sky patches centred at cluster positions allows us to get the best estimates of the local noise properties. The MMF also makes assumptions about the spatial and spectral characteristics of the cluster signal and the instrument.  Our cluster model is described below, for the instrumental response we assumed symmetric Gaussian beams with FWHM given in Table~\ref{tab:fwhm}.
  
We determine a single quantity for each cluster from the \planck\ data, the normalisation of an assumed profile. All the parameters determining the profile location, shape and size are fixed using X-ray data. We use the profile shape described in Sect. \ref{clustermodel.sec} with $c_{500}$,  $\alpha$, $\beta$ and $\gamma$ fixed to the values given in Table \ref{tab:pressure_param} and integrate along the line-of-sight to obtain a template for the cluster SZ signal. The integral is performed by considering a cylindrical volume and a cluster extent of $5 \times R_{\rm 500}$ along the line-of-sight. The exact choice of the latter is not relevant. The normalisation of the profile is fitted using data within a circular aperture of radius $5 \times R_{\rm 500}$ for each system in our X-ray cluster catalogue, centring the filter on the X-ray position and fixing the cluster size to $\theta_{500} = R_{500}/D_{\rm A}(z)$. Notice that the dependence of cluster size on X-ray luminosity is weak ($R_{\rm 500} \propto L_{\rm 500}^{0.2}$ from Eq. \ref{L -- M:eq}), implying that MMF measurements are expected to be relatively insensitive to the details of the underlying $L_{\rm 500}$ -- $M_{\rm 500}$ relation.
\begin{figure*}[ht!]
 \begin{centering}
\includegraphics[width=1. \columnwidth]{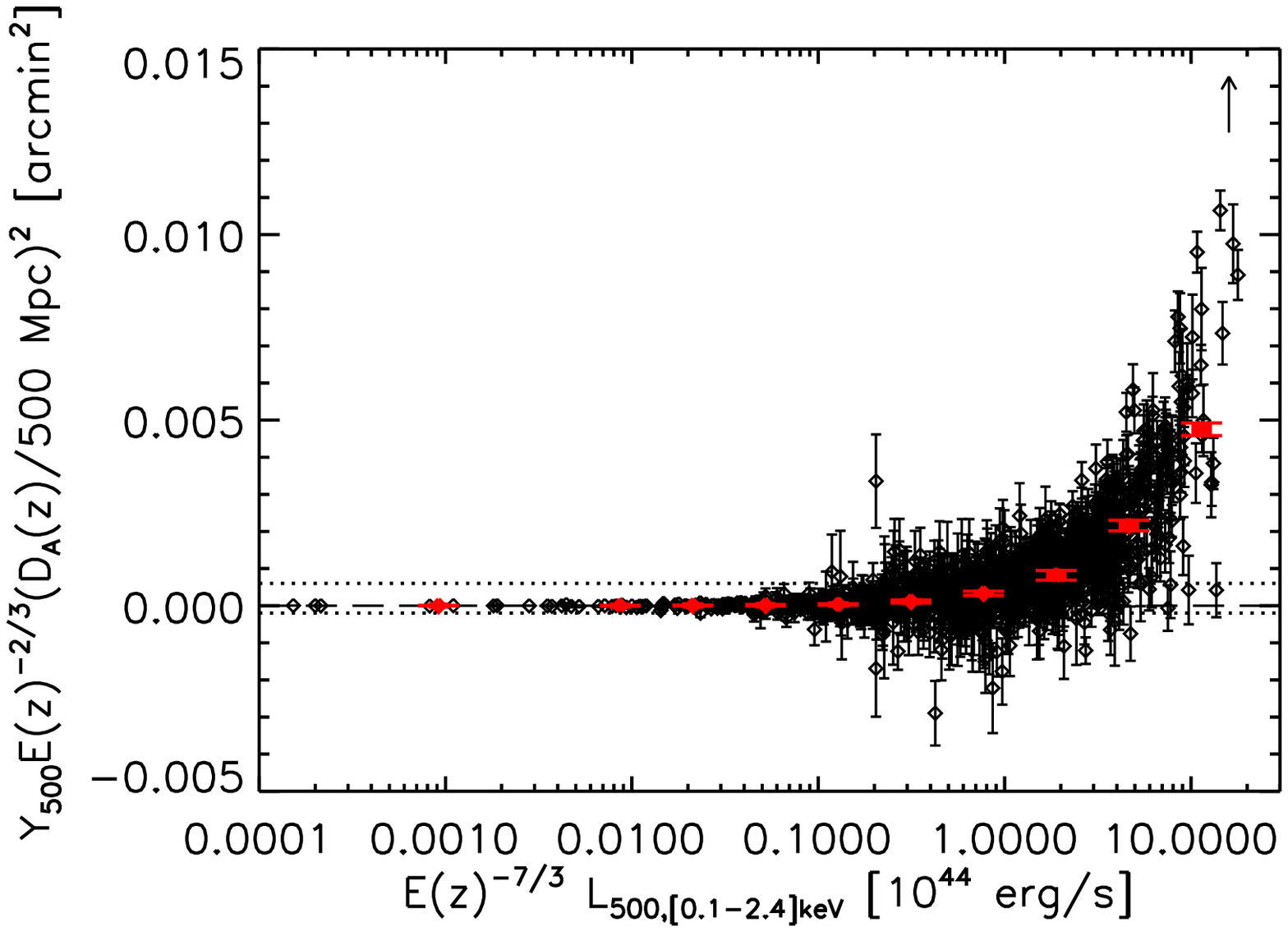}
\hfill
\includegraphics[width=1. \columnwidth]{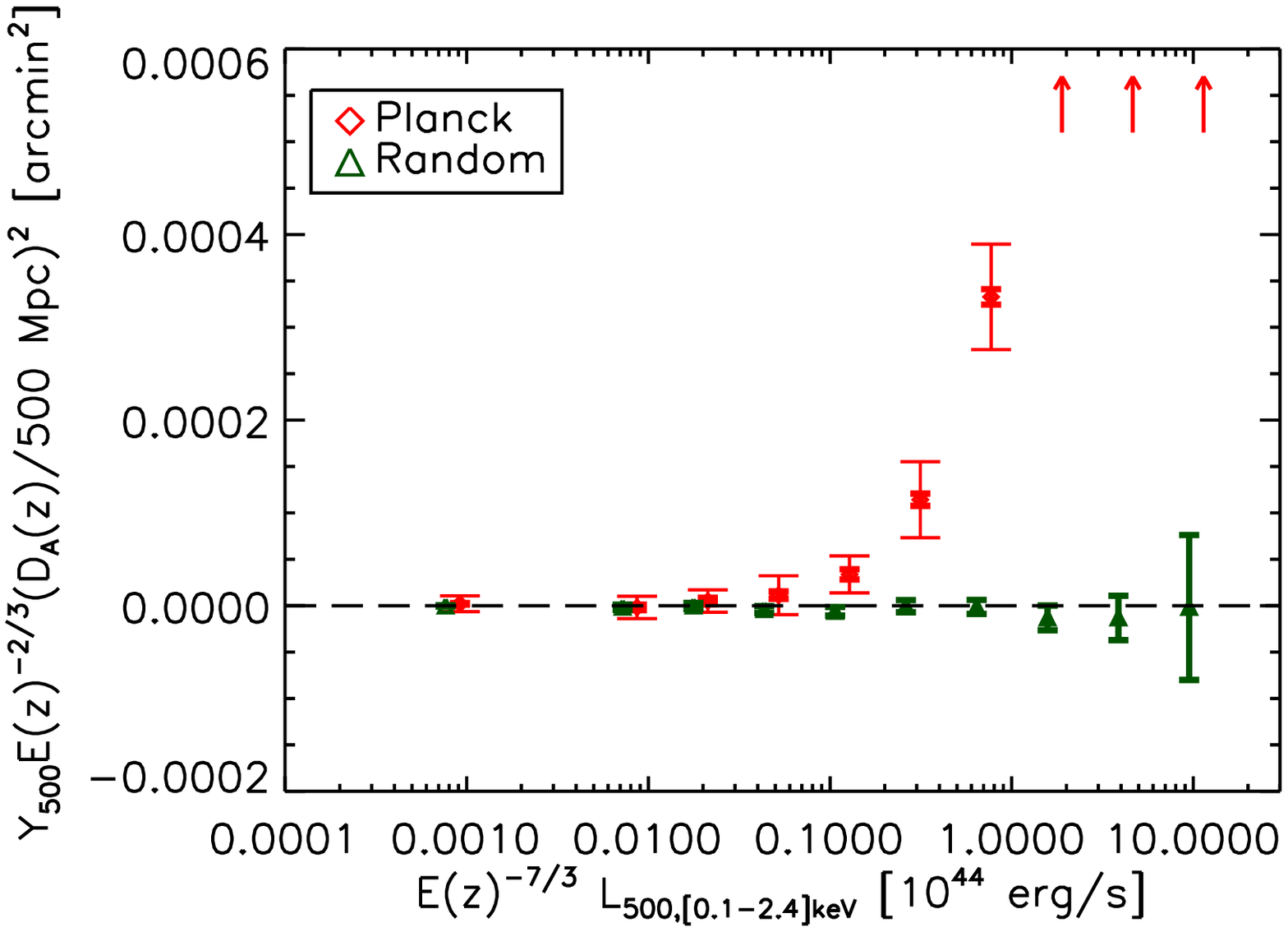}
\end{centering}

 \caption{{\footnotesize {\it Left:\/} Intrinsic SZ signal from a sphere of radius $R_{500}$ as a function of the X-ray luminosity for all the clusters in the sample individually. Error bars indicate the pure measurement uncertainties based on MMF noise estimates (statistical uncertainties). Red diamonds show the bin averaged values with thick and thin error bars indicating the statistical (not visible) and bootstrap uncertainties, respectively.  {\it Right:\/} Zoom onto the scale indicated by the horizontal dotted lines in the left-hand panel. Red symbols and error bars as in left-hand panel. Green triangles (shifted towards lower X-ray luminosity values by 20\% with respect to diamonds for clarity) show the result of the same analysis when the signal is estimated at random positions instead of true cluster positions. The associated thick error bars indicate the statistical uncertainties.}}
    \label{y5r500:fig}
  \end{figure*}

The MMF method yields statistical SZ measurement errors $\sigma_{\rm i}$ on individual meaurements. The statistical error includes uncertainties due to the instrument (beam, noise) and to the astrophysical contaminants (primary CMB, Galaxy, point sources). Obviously, it does not take into account the uncertainties on our X-ray priors and instrumental properties which will be studied in Sect. \ref{robustness.sec}  

The same extraction method is used in \cite{planck2011-5.2c} where the optical--SZ scaling relation with MaxBCG clusters \citep{koester07} are investigated. There are only two differences. First, in this paper we use the X-ray scaling $L_{\rm 500}$ -- $M_{\rm500}$ to adapt our filters to the sizes of our clusters while we use the optical $N_{\rm 200}$ -- $M_{\rm 500}$ relation of \cite{johnston} and \cite{rozo} in the other paper. Second, the MaxBCG catalogue includes $\sim$ 14,000 clusters so we do not build a set of patches for each cluster individually. Instead, we divide the sphere into 504 overlapping patches  ($10^\circ \times 10^\circ$, pixel=1.72 arcmin) as in \cite{melin2010}. We also extract SZ signal for each cluster individually but the clusters are no longer located at the centre of the patch.

Under the assumption that the shape of the adopted profile template corresponds to the true SZ signal, our extraction method allows us to convert the signal in a cylinder of aperture radius $5 \times R_{\rm 500}$ to $Y_{\rm 500}$, the SZ signal in a sphere of radius $R_{\rm 500}$. By definition the conversion factor is a constant factor for every cluster but depends on the assumed profile. The effect of the uncertainties on the assumed profile are discussed in Sect. \ref{robustness.sec} below. 

The {\it intrinsic SZ signal\/} is computed by taking into account the angular distance dependence of the observed signal and is expressed 
as $( {D_{\rm A}(z) / 500 \, {\rm Mpc}}  )^{2} \, Y_{\rm 500}$. This signal has units of ${\rm arcmin}^2$ as for the observed quantity, but its value differs from the intrinsic signal in units of ${\rm Mpc}^2$ by a constant, redshift-independent factor. Making such a conversion allows us to directly compare our measurements with the predictions derived from the model detailed in Sect. \ref{clustermodel.sec} (see in particular  Eq.~\ref{YM} and~\ref{YM_arc}). 
When a specific scaling relation is investigated, the SZ signal is appropriately scaled according to the adopted scaling relations presented in Sect. \ref{clustermodel.sec} (e.g., see the left-hand panel of Fig. \ref{y5r500:fig}). 

The SZ signal for all the clusters in our sample is shown as a function of the [0.1-2.4] keV band X-ray luminosity in the left-hand panel of Fig. \ref{y5r500:fig}. Assuming standard evolution the intrinsic quantities $( {D_{\rm A}(z) / 500 \, {\rm Mpc}} )^{2} \, Y_{\rm 500} \, E(z)^{-2/3}$ are plotted as a function of  $L_{\rm 500} E(z)^{-7/3}$. The figure shows that \planck \ detects the SZ signal at high significance for a large fraction of the clusters.

\subsection{Binned SZ signal}
\label{binnedfluxes}

As shown in the left-hand panel of Fig. \ref{y5r500:fig} the SZ signal is not measured at high significance for all of the clusters. In particular, low luminosity objects are barely detected individually. We therefore take advantage of the large size of our sample and average SZ measurements in X-ray luminosity, mass, or redshift bins. The bin average of the intrinsic SZ signal is defined as the weighted mean of the signal in the bin (with inverse variance weight, $\sigma_i^{-2}$, scaled to the appropriate redshift or mass dependence depending on the studied scaling relation) and the associated statistical errors are computed accordingly. The binning depends on the adopted relation and will be detailed in each case. 

In the left-hand panel of Fig. \ref{y5r500:fig} the binned signal is overlaid on the individual measurements. In this case the SZ signal is averaged in logarithmically spaced luminosity bins. We merged the lowest four luminosity bins into a single bin to obtain a significant result. The statistical uncertainties, which are depicted by the thick error bars, are not visible in the figure and clearly underestimate the uncertainty on the binned values. 

A better estimate of the uncertainties in the binned values comes from an ensemble of 10,000 bootstrap realisations of the entire X-ray cluster catalogue. Each realisation is constructed from the original data set by random sampling with replacement, where all quantities of a given cluster are replaced by those of another cluster. Each realisation is analysed in the same way as the original catalogue and the standard deviation of the average signal in each bin is adopted as total error. Bootstrap uncertainties, which take into account both sampling and statistical uncertainties, are shown by the thin error bars in the left-hand panel of Fig. \ref{y5r500:fig}. A visual inspection of the figure indicates that the SZ signal is detected at high significance over a wide luminosity range. The lack of clear detection at $L_{\rm 500} E(z)^{-7/3} \lesssim 0.05 \times 10^{44} {\rm erg/s}$ is due to the combined effect of low signal and small number of objects. In the companion paper \cite{planck2011-5.2c} we explore this low luminosity (mass) range in more depth. The results of these two complementary analysis are summarised and discussed in Sec. \ref{conclusions.sec} (see Fig. \ref{optical_comp:fig} and related discussion). 

\begin{figure}
  \includegraphics[width=1.000 \columnwidth]{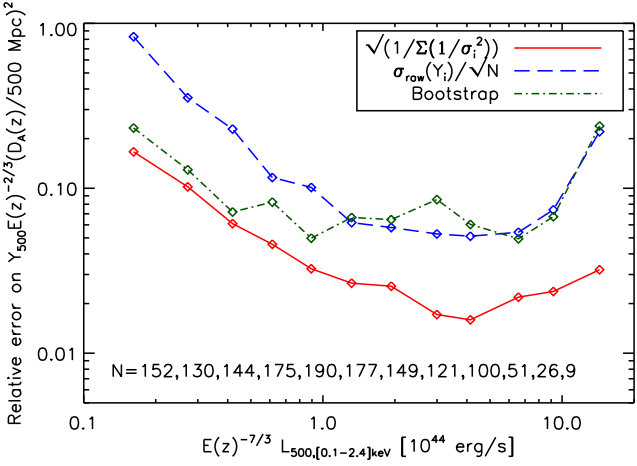}
   \caption{{\footnotesize Bin averaged relative statistical errors (solid line) and relative bootstrap errors (dot-dashed line) are shown as a function of X-ray luminosity. The numbers given in the legend indicate the number of objects in each luminosity bin. For comparison, the scaled unweighted standard deviation (dashed line) is also shown.}}
       \label{errors:fig}
  \end{figure}

The difference between statistical and bootstrap errors are rendered in more detail in Fig. \ref{errors:fig}, where relative bootstrap uncertainties (dot-dashed line) are compared to in-bin relative statistical errors (solid line). The figure shows that for $L_{\rm 500} E(z)^{-7/3} \lesssim 10^{44} {\rm erg/s}$ statistical uncertainties are dominant. This implies that intrinsic scatter, which is discussed in more detail in Sect.\ref{scatter.sec}, can only be measured at higher luminosity. 

Fig. \ref{errors:fig} also shows the quantity $(1 / \sqrt{N}) \times (\sigma_{\rm raw} / Y)$ (dashed line), which is computed from the unweighted raw scatter $\sigma_{\rm raw}$, the bin average $Y$, and the number of clusters in the bin N. The difference between the latter and the relative bootstrap uncertainties in the low luminosity bins is due to the range of relative errors on individual measurements in a given bin. 

As a robustness check, we have undertaken the analysis a second time using random cluster positions but keeping all the properties of our sample (sizes, profile shape). The result is shown by the green triangles in the right-hand panel of Fig. \ref{y5r500:fig} and, as expected, is consistent with no detection of the SZ signal. This demonstrative {\it null test\/} clearly shows the efficiency of the MMF to pull out the SZ signal from our cluster sample. Additional robustness test are discussed in Sect. \ref{robustness.sec} below.
\section{Results}
\label{scalingrelations.sec}

\subsection{The $D^2_{\rm A} \, Y_{\rm 500}$ -- $L_{\rm 500}$ and $D^2_{\rm A} \, Y_{\rm 500}$ -- $M_{\rm 500}$ relations} 
\label{y-l.sec}

\begin{figure*}[!ht]
 \begin{centering} 
\includegraphics[width=1. \columnwidth]{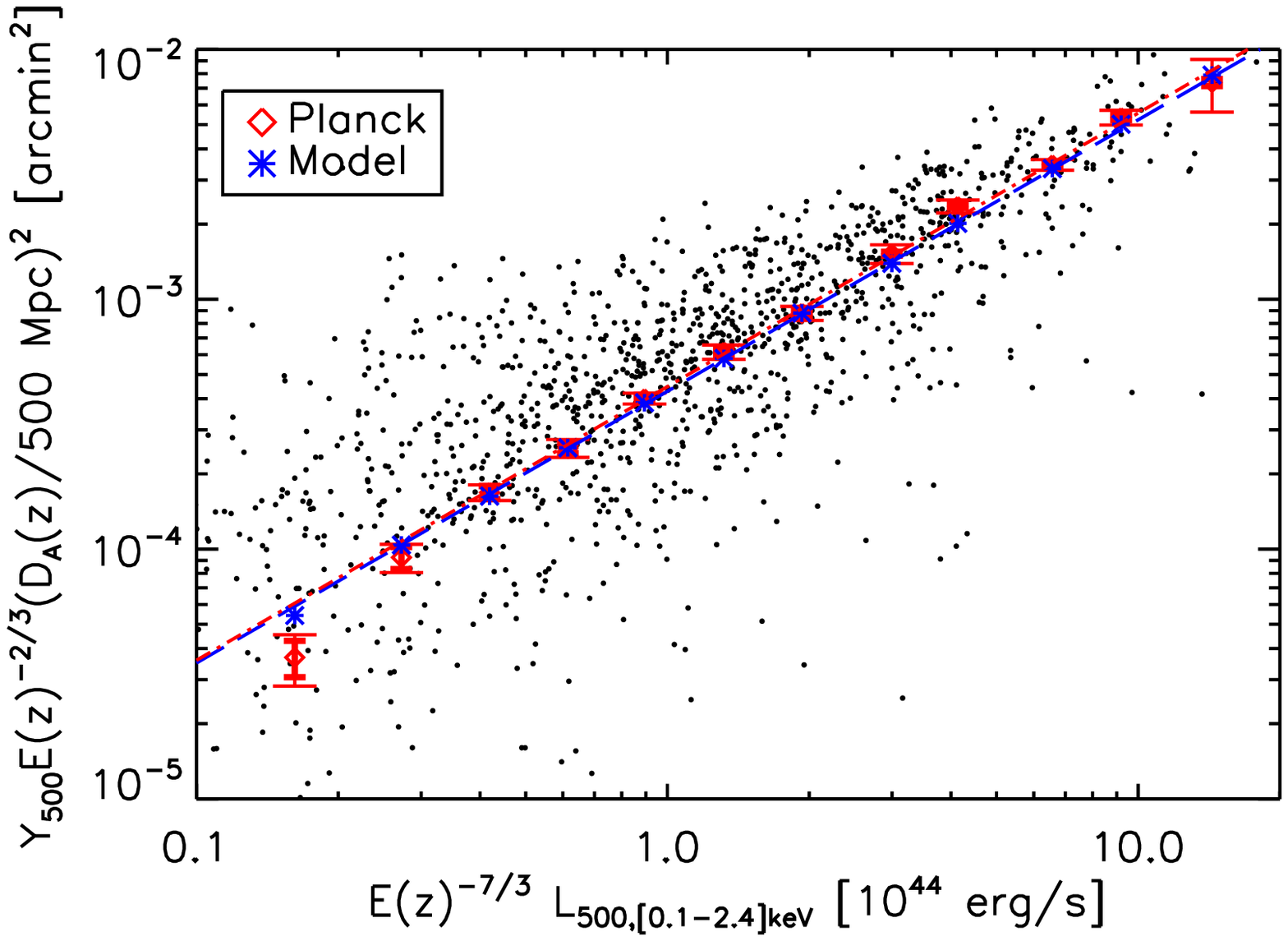}
\hfill 
\includegraphics[width=1. \columnwidth]{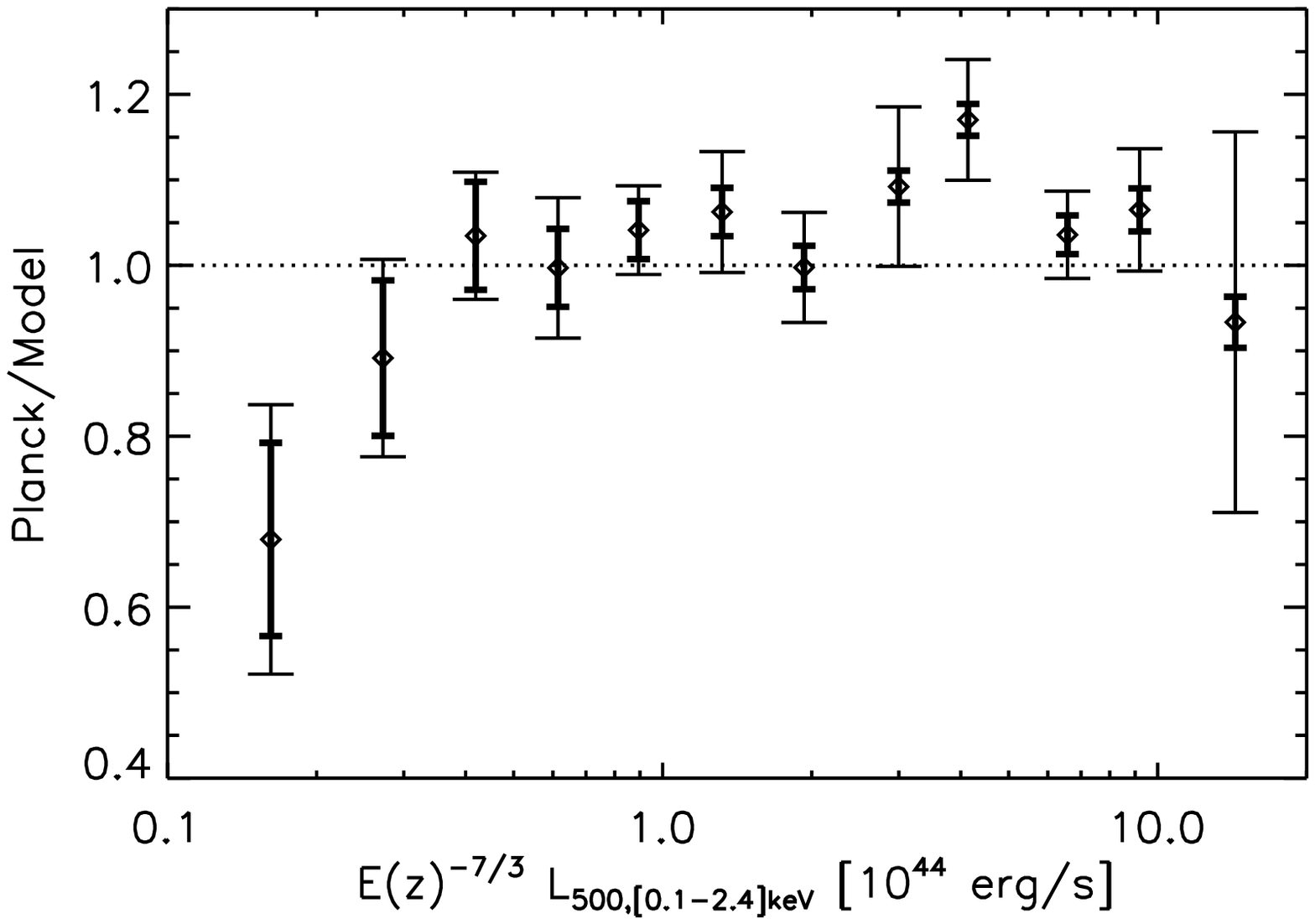} 
\end{centering} 
 \caption{{\footnotesize {\it Left:\/} Scaling relation between \planck \ SZ measurements and X-ray luminosity for $\sim$ 1600 MCXC clusters. Individual measurements are shown by the black dots and the corresponding bin averaged values by the red diamonds. Thick bars give the statistical errors, while the thin bars are bootstrap uncertainties. The bin-averaged SZ cluster signal expected from the X-ray based model is shown by the blue stars. The combination of the adopted $D^2_{\rm A} \, Y_{\rm 500}$ -- $M_{\rm 500}$ and $L_{\rm 500}$ -- $M_{\rm 500}$ relations (Eq. \ref{YL}) is shown by the dashed blue line while the red dot-dashed line shows the best fitting power-law to the data (Eq. \ref{YLfit} and Table \ref{fit_param_un_l}). {\it Right:\/} Ratio between data and model bin averaged values shown in the left panel. Error bars are as in the left panel.}}
    \label{planckvsmodel:fig} 
  \end{figure*}

The main results of our analysis are summarised in Fig. \ref{planckvsmodel:fig}. In the left-hand panel of the figure the individual and luminosity binned \planck \ SZ signal measured at the location of MCXC clusters are shown as a function of luminosity together with the luminosity averaged model predictions. The latter are computed by averaging the model prediction for individual clusters (see Sect. \ref{clustermodel.sec}) with the same weights as for the measured signal. Notice that SZ signal and X-ray luminosity are intrinsic quantities and are scaled assuming standard evolution. The figure shows the high significance of the SZ signal detection and the excellent agreement between measurements and model predictions. 
The agreement between \planck \ measurements and X-ray based predictions is rendered in more detail in the right-hand panel of Fig. \ref{planckvsmodel:fig} where the \planck -to-model ratio is plotted. Taking into account the total errors given by the bootstrap uncertainties (thin bars in the figure), the agreement is excellent over a wide luminosity range.
We model the observed $D^2_{\rm A} \, Y_{\rm 500}$ -- $L_{\rm 500}$ relation shown in the left-hand panel of Fig. \ref{planckvsmodel:fig} by adopting a power law of the form
\begin{equation} 
\label{YLfit}
Y_{500} = \hat{Y}_{500,L} \; \left ({E(z)^{-7/3} L_{500} \over 10^{44} {\rm erg/s}} \right )^{\hat{\alpha}_{L}} \; E(z)^{\hat{\beta}_{L}} \; \left ({D_{A}(z) \over 500 \, {\rm Mpc}} \right )^{-2} 
\end{equation} 
and directly fitting the individual points shown in the figure rather than the binned data points. We use a non-linear least-squares fit built on a gradient-expansion algorithm (the IDL curvefit function). In the fitting procedure, only the statistical errors given by the MMF are taken into account. The derived uncertainties on the best fitting parameters are quoted in Table \ref{fit_param_un_l} as statistical errors. In addition, uncertainties on the best fitting parameters are estimated through the bootstrap procedure described in Sec.~\ref{SZextraction.sec}. Each bootstrap catalogue fit leads to a set of parameters whose standard deviation is quoted as the uncertainty on the best fitting parameters. Values are given for three different choices of priors as given in Table \ref{fit_param_un_l}, where the best fitting parameters are listed. The table also provides the prediction of our X-ray based model for comparison.

Fixing the slope and the redshift dependence of the $D^2_{\rm A} \, Y_{\rm 500}$ -- $L_{\rm 500}$ relation, the best fitting amplitude is $0.451 \times 10^{-3} \, {\rm arcmin^2}$, in agreement with the model prediction $0.428 \times 10^{-3} \, {\rm arcmin^2}$ at $1.8 \sigma$. When keeping the redshift dependence of the relation fixed but leaving the slope of the relation free, we find agreement between best fitting and predicted slopes at better than $1 \sigma$, while the amplitudes remain in agreement at $1.3 \sigma$. For maximum usefulness and in particular to facilitate precise comparisons with our findings, we provide, in Table \ref{binnedvalues}, the data points shown in the left-hand panel of Fig. \ref{planckvsmodel:fig}. Values are given for the quantities $\tilde{L}_{\rm 500}= L_{\rm 500} E(z)^{-7/3} $ in units of $10^{44} {\rm erg/s}$ and  $\tilde{Y}_{\rm 500}=Y_{\rm 500} E(z)^{-2/3} \left ( {D_{\rm A}(z) / 500 \, {\rm Mpc}} \right )^{2}$ in units of $10^{-3} \, {\rm arcmin}^2 $.  Both total (i.e., bootstrap) and statistical errors on $\tilde{Y}_{\rm 500}$ are also listed.
    \begin{table*} 
     \caption[]{{\footnotesize Best fitting parameters for the observed $D^2_{\rm A} \, Y_{\rm 500}$ -- $L_{\rm 500}$ relation given in Eq. \ref{YLfit}. Values are given for three different choices of priors and as predicted from X-rays for comparison. Both total errors from bootstrap resampling and statistical errors are quoted.}}

        \label{fit_param_un_l} 
       \begin{center} 
       \begin{tabular}{lccc} 
        \hline 
        \hline 
       & $\hat{Y}_{\rm 500, L}$ [$10^{-3} \, {\rm arcmin}^2 $] & $\hat{\alpha}_{\rm L}$ & $\hat{\beta}_{\rm L}$ \\ 
         \hline 
         &   $ 0.451 \pm 0.003 \, {\rm stat} \; [\pm 0.013 \, {\rm tot} ]  $ & 1.087 (fixed) & 2/3 (fixed) \\

     \planck \ + MCXC & $ 0.447 \pm 0.006 \, {\rm stat} \; [ \pm 0.015 \, {\rm tot} ] $ & $1.095 \pm 0.008 \, {\rm stat} \; [ \pm 0.025 \, {\rm tot}]$ & 2/3 (fixed) \\
     
           &  $ 0.476 \pm 0.006 \, {\rm stat} \; [ \pm 0.025 \, {\rm tot} ] $ & 1.087 (fixed) &  $-0.007 \pm 0.154 \, {\rm stat} \; [ \pm 0.518 \, {\rm tot} ]$\\

         \hline 
                   
         X-ray prediction     &     $ 0.428 $ & 1.09 &  2/3 \\

         \hline 
                                     
      \end{tabular} 
     \end{center} 
  \end{table*} 
    \begin{table} 
     \caption[]{{\footnotesize Bin averages of the $D^2_{\rm A} \, Y_{\rm 500}$ -- $L_{\rm 500}$ relation shown in the left panel of Fig. \ref{planckvsmodel:fig}.}}

        \label{binnedvalues} 
       \begin{center} 
       \begin{tabular}{cccccc} 
        \hline 
        \hline 
                              
             $\tilde{L}_{\rm 500}$  & Nr.  & $\tilde{L}_{\rm 500}$  & $\tilde{Y}_{\rm 500}$ & $\Delta \tilde{Y}_{\rm 500}$ & $\Delta \tilde{Y}_{\rm 500}$ \\         
range & Obj. & &  & statistical & total \\ 
             \hline

0.100 - 0.222 & 152 & 0.162 & 0.037 & 0.006 & 0.009 \\
0.222 - 0.331 & 130 & 0.272 & 0.093 & 0.009 & 0.012 \\
0.331 - 0.493 & 144 & 0.419 & 0.169 & 0.010 & 0.012 \\
0.493 - 0.734 & 175 & 0.615 & 0.254 & 0.012 & 0.021 \\
0.734 - 1.094 & 190 & 0.894 & 0.401 & 0.013 & 0.020 \\
1.094 - 1.630 & 177 & 1.319 & 0.616 & 0.016 & 0.041 \\
1.630 - 2.429 & 149 & 1.931 & 0.879 & 0.022 & 0.057 \\
2.429 - 3.620 & 121 & 2.997 & 1.521 & 0.026 & 0.130 \\
3.620 - 5.393 & 100 & 4.138 & 2.356 & 0.038 & 0.142 \\
5.393 - 8.036 & 51 & 6.572 & 3.456 & 0.076 & 0.171 \\
8.036 - 11.973 & 26 & 9.196 & 5.342 & 0.126 & 0.359 \\
11.973 - 17.840 & 9 & 14.345 & 7.369 & 0.236 & 1.758 \\

         \hline 
                                     
      \end{tabular} 
     \end{center} 
  \end{table} 

For completeness, we also investigate the $D^2_{\rm A} \, Y_{\rm 500}$ -- $M_{\rm 500}$ relation, where the masses $M_{\rm 500}$ are computed from the $L_{\rm 500}$ -- $M_{\rm 500}$ relation given in Eq. \ref{L -- M:eq}. Following the same procedure as for the $D^2_{\rm A} \, Y_{\rm 500}$ -- $L_{\rm 500}$ relation, we fit individual points of the $D^2_{\rm A} \, Y_{\rm 500}$ -- $M_{\rm 500}$ plane with
\begin{equation} 
\label{YMfit}
 Y_{500} = \hat{Y}_{500,M} \; \left ({M_{500} \over 3 \times 10^{14}  M_\odot} \right )^{\hat{\alpha}_{M}} \; E(z)^{\hat{\beta}_{M}} \; \left ({D_{A}(z) \over 500 \, {\rm Mpc}} \right )^{-2}.
 \end{equation}
 The same cases as for the $D^2_{\rm A} \, Y_{\rm 500}$ -- $L_{\rm 500}$ relation are considered and the best fitting parameters are provided in Table \ref{fit_param_un} along with the model prediction. Concerning the agreement between best fitting parameters and model predictions, the conclusions drawn for the $D^2_{\rm A} \, Y_{\rm 500}$ -- $L_{\rm 500}$  relation obviously apply also for the $D^2_{\rm A} \, Y_{\rm 500}$ -- $M_{\rm 500}$.
   \begin{table*} 
     \caption[]{{\footnotesize Best fitting parameters for the observed $D^2_{\rm A} \, Y_{\rm 500}$ -- $M_{\rm 500}$ relation given in Eq. \ref{YMfit}. Values are given for three different choices of priors and as predicted from X-rays for comparison.  Both total errors from bootstrap resampling and statistical errors are quoted.}}

        \label{fit_param_un} 
       \begin{center} 
       \begin{tabular}{cccc} 
        \hline 
        \hline 
     &    $\hat{Y}_{\rm 500, M}$ [$10^{-3} \, {\rm arcmin}^2$] & $\hat{\alpha}_{\rm M}$ & $\hat{\beta}_{\rm M}$ \\ 
         \hline 
       &        $ 0.896 \pm 0.007 \, {\rm stat} \; [ \pm 0.027 \, {\rm tot} ] $ & 1.783 (fixed) & 2/3 (fixed) \\

 \planck \ + MCXC  & $ 0.892 \pm 0.008 \, {\rm stat} \; [ \pm 0.025 \, {\rm tot} ] $ & $1.796 \pm 0.014 \, {\rm stat} \; [ \pm 0.042 \, {\rm tot}]$ & 2/3 (fixed) \\

         &    $ 0.945 \pm 0.012 \, {\rm stat} \; [ \pm 0.049 \, {\rm tot} ] $ & 1.783 (fixed) &  $-0.007 \pm 0.154 \, {\rm stat} \; [ \pm 0.518 \, {\rm tot}]$\\
      
         \hline 
                   
         X-ray prediction     &     $ 0.850 $ & 1.783 &  2/3 \\

        \hline 
     \end{tabular} 
       \end{center} 
  \end{table*} 

\subsection{Redshift evolution}
\label{z_evol.sec}
\begin{figure}
 \begin{centering}
\includegraphics[width=1.000 \columnwidth]{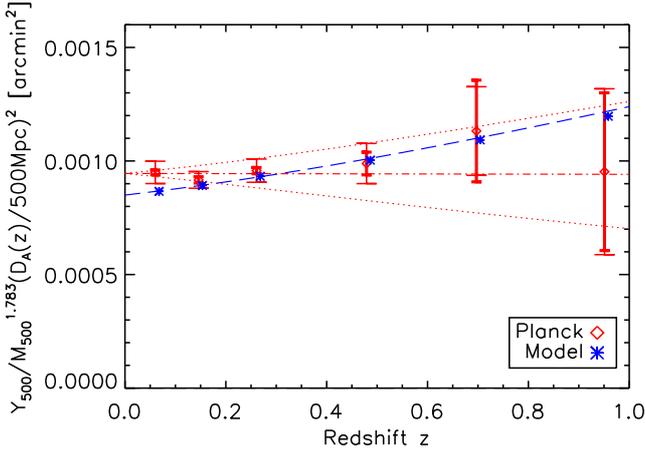}
\end{centering}

\caption{{\footnotesize Bin averaged SZ signal from a sphere of radius $R_{500}$ ($Y_{500}$) scaled by the expected mass and angular distance dependence as a function of redshift.  The \planck \ data (red diamonds) and the SZ cluster signal expected from the X-ray based model  (blue stars) are shown together with the expected standard redshift evolution (dahed line). The best fitting model is shown by the dot-dashed line and the $1 \sigma$ confidence region is limited by the dotted lines. Here $M_{\rm 500}$ is given in units of  $ 3 \times 10^{14}  M_\odot$.}}
    \label{y500z:fig}

  \end{figure}

We also considered the case where the redshift evolution of the scaling relations is allowed to differ from the standard expectation.
Using the simplest model (Eq. \ref{YLfit} or equivalently Eq. \ref{YMfit}) we attempt to constrain 
the power law index $\hat{\beta}_{\rm L}$ (or equivalently $\hat{\beta}_{\rm M}$).
We find that the measured SZ signal is consistent with standard evolution (see Table \ref{fit_param_un_l}) and
our constraints on any evolution are weak.
Fig. \ref{y500z:fig} shows the measured and predicted, redshift binned, SZ signal, the expected standard redshift evolution, and the best fitting model.
The figure shows that, although measurements and predictions agree quite well, the best fitting model is constrained primarily by the low redshift measurements. Possible future improvements are discussed below in Sect. \ref{conclusions.sec}.

\subsection{Scatter in the $D^2_{\rm A} \, Y_{\rm 500}$ -- $L_{\rm 500}$ relation}
\label{scatter.sec}

As discussed in Sect. \ref{binnedfluxes}, we find a clear indication of intrinsic scatter in our measurements of the $D^2_{\rm A} \, Y_{\rm 500}$ -- $L_{\rm 500}$ relation. In this section we quantify this scatter and discuss how our measurement compares with expectations based on the representative \rexcess \ sample \citep{arnaud2010} and the findings reported in the companion paper discussing high quality observations of local clusters \citep{planck2011-5.2b}.

The intrinsic scatter $\sigma_{\rm intr}$ is computed in luminosity bins as the quadratic difference between the raw scatter $\sigma_{\rm raw}$ (see Sect. \ref{binnedfluxes}) and the statistical scatter expected from the statistical uncertainties, i.e. $\sigma^2_{\rm intr}=\sigma^2_{\rm raw}-\sigma^2_{\rm stat}$. The latter is estimated by averaging the statistical uncertainties in a given bin, i.e. $\sigma^2_{\rm stat}=N^{\rm -1} \, \sum \sigma^2_{\rm i}$, where N is the number of clusters in the bin. For a given luminosity bin, the uncertainty $\Delta \sigma_{\rm intr}$ on the estimated intrinsic scatter are evaluated by $(\Delta \sigma_{\rm intr})^2 = \sigma_{\rm intr}^2 (2 \, N \, (N-1))^{-1} \sum (1 + (\sigma_i^2/\sigma_{\rm intr}^2))^2 $.

\begin{figure}
  \includegraphics[width=1.000 \columnwidth]{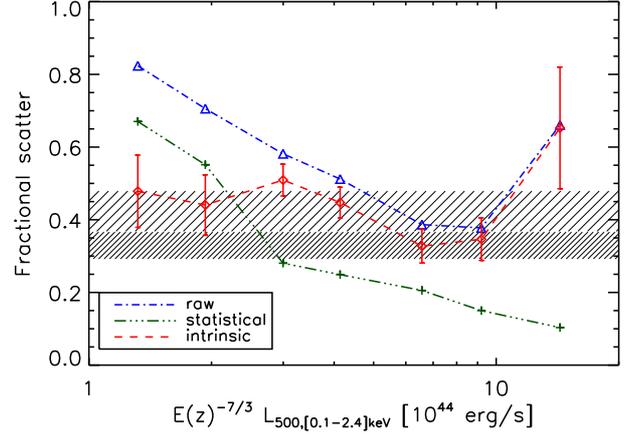}
     \caption{{\footnotesize Fractional raw (dot-dashed blue line and triangles), statistical (dot-dot-dashed green line and plus signs), and intrinsic (dashed red line, diamonds, and error bars) scatter on the $D^2_{\rm A} \, Y_{\rm 500}$ -- $L_{\rm 500}$ relation.  The coarse/fine-hatched regions corresponds to the 1 $\sigma$ uncertainties on the intrinsic scatter reported in \cite{arnaud2010} and \cite{planck2011-5.2b}, respectively.}}
       \label{scatter:fig}
  \end{figure}

\begin{figure}
  \includegraphics[width=1.000 \columnwidth]{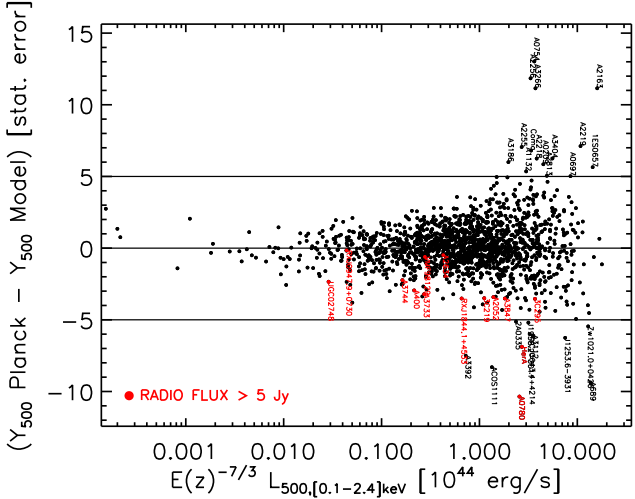}
  \caption{{\footnotesize Difference between the \planck \ measurement and the X-ray based prediction in units of the measurement statistical error $\sigma_{\rm i}$ (pure measurement uncertainties based on MMF noise estimates) as a function of X-ray luminosity. Labelled black points denote objects with a difference larger than 5 $\sigma_{\rm i}$ and are further discussed in the text. Clusters with SZ signal possibly contaminated by radio sources (see discussion in Sect. \ref{robustness.sec}) are shown in red and labelled by their name.}}
  \label{planck_model:fig}
\end{figure}

We find that intrinsic scatter can be measured only for $L_{\rm 500} E(z)^{-7/3} \gtrsim 10^{44} {\rm erg/s}$, because the statistical uncertainties at lower luminosities are close to the value of the raw scatter (see also Sect. \ref{binnedfluxes}). In a given bin with average signal $Y$, the resulting fractional intrinsic scatter $\sigma_{\rm intr}/Y$ is shown in Fig. \ref{scatter:fig} along with the fractional raw and statistical scatters. The estimated intrinsic scatter is close to $40-50 \%$ and in agreement with the expectations given in \cite{arnaud2010} ($\sigma_{\rm log Y_{\rm 500}} = 0.184 \pm 0.024$, the range of these values is indicated by the coarse--hatched region in the figure). Notice that the intrinsic scatter reported in \cite{arnaud2010} is computed for the \rexcess \ sample and evaluated adopting {\it XMM-Newton\/} luminosities and a predicted SZ signal for individual objects based on the same model assumed here but relying on the mass proxy $Y_{\rm X}$. Therefore, the intrinsic scatter quoted in  \cite{arnaud2010} reflects the intrinsic scatter in the underlying $L_{\rm 500}$ -- $M_{\rm 500}$ relation. In \cite{planck2011-5.2b}, where a sample of clusters detected at high signal to noise in the \planck \ survey \citep[the ESZ sample, see][]{planck2011-5.1a} and with high quality X-ray data from {\it XMM-Newton\/} is used, the intrinsic scatter in the $D^2_{\rm A} \, Y_{\rm 500}$ -- $L_{\rm 500}$ relation is found to be $\sigma_{\rm log Y_{\rm 500}} = 0.143 \pm 0.016$. These values are shown in Fig. \ref{scatter:fig} by the fine--hatched region. In  \cite{planck2011-5.2b} it is found that cool core clusters are responsible for the vast majority of the scatter around the relation. Because the sample used in this study is X-ray selected, we expect it to contain a higher fraction of cool core systems than in the ESZ subsample studied in \cite{planck2011-5.2b}. This implies that the scatter in the $D^2_{\rm A} \, Y_{\rm 500}$ -- $L_{\rm 500}$ relation measured in our sample is expected to be higher than the one found in \cite{planck2011-5.2b}, as observed.
Given the segregation of cool core systems in the $D^2_{\rm A} \, Y_{\rm 500}$ -- $L_{\rm 500}$ reported in \cite{planck2011-5.2b}, we investigate the link between the intrinsic scatter in the relation and cluster dynamical state using our large X-ray sample. To this end we compare  \planck \ measurements and the X-ray based predictions (i.e., Eq. \ref{YL}) for individual objects. In Fig. \ref{planck_model:fig} we show the difference between \planck \ measurement and the X-ray based prediction in units of the measurement statistical error $\sigma_{\rm i}$ (see Sect. \ref{SZextraction.sec}) as a function of X-ray luminosity and investigate the largest outliers the figure, i.e. the most statistically significant outliers in the $D^2_{\rm A} \, Y_{\rm 500}$ -- $L_{\rm 500}$ relation.  

Given the size of our sample we discuss individually only clusters for which measurement and prediction differ by more than 5 $\sigma_{\rm i}$.  These are identified and information on their dynamical state is searched for in the literature. Information is based on the classification of \cite{hudson2010} if not stated otherwise. 
We find 15 clusters with a predicted signal smaller than the \planck \ measurement by more than 5 $\sigma_{\rm i}$. Of these, six are known merging clusters: Coma, A2218 \citep{govoni2004}, 1ES0657, A754 \citep{govoni2004}, A2163 \citep{bourdin2010}, A0697 \citep{girardi2006}, six are classified as non-cool core clusters and may therefore be unrelaxed: A2219 \citep{allen1998}, A2256, A2255, A0209 \citep{zhang2008}, A2813 \citep{zhang2008}, A3404 \citep{pratt2009}, and A3266 is a weak cool core cluster. No information is available for the remaining clusters: A1132 and A3186. Conversely, there are 11 over-predicted clusters at 5 $\sigma_{\rm i}$. Of these five are strong cool core clusters: 2A0335, Zw1021.0+0426 \citep{morandi2007}, A3112, HerA \citep{bauer2005} , and A0780. No information is available for the remaining clusters: A689, ACOS1111, A3392, J1253.6-3931, J1958.2-3011, and RXCJ0643.4+4214. Notice that the luminosity of A689 is likely to be overestimated by a large factor because of point source contamination \citep{maughan2008}. In addition, for A3186 and ACOS1111 the model prediction rely on the EMSS luminosity measurements given in \cite{emss1994}, which might be unreliable. 

The high fraction of dynamically perturbed / cool-core clusters with largely under/over predicted SZ signal is confirmed when additional outliers at smaller $\sigma_{\rm i}$ are searched. We find 65 clusters with $(Y_{\rm 500} \, {\rm Planck} - Y_{\rm 500} \,  {\rm Model} ) > \, 3 \, \sigma_{\rm i}$ and that for 46 percent of them dynamical state information is available in the literature. Of the latter 26 (87 percent) are either known mergers or non-cool core clusters, two are weak cool core clusters, and only one is classified as cool core cluster. We find 53 clusters with $(Y_{\rm 500} \, {\rm Planck} - Y_{\rm 500} \,  {\rm Model} ) < \, -3 \, \sigma_{\rm i}$. For 45 percent of these clusters we are able to find information on their dynamical state in the literature and find that 96 percent of them are cool core clusters. These findings clearly suggest that the intrinsic scatter in the $D^2_{\rm A} \, Y_{\rm 500}$ -- $L_{\rm 500}$ relation is linked to the cluster dynamical state, as also found in \cite{planck2011-5.2b}.

\section{Robustness of the results}
\label{robustness.sec}

As in the other four \planck\ SZ papers \citep{planck2011-5.1a,planck2011-5.1b,planck2011-5.2b,planck2011-5.2c}, we test the robustness of our results for the effect of several instrumental, modelling, and astrophysical uncertainties. Tests common to all \planck\ SZ papers are discussed in detail in Sec.~6 of \citet{planck2011-5.1a}. Of these, calibration and colour correction effects are relevant for our analysis. Calibration uncertainties are shown to propagate into very small uncertainties on SZ signal measurements ($\sim 2 \%$) and colour correction is found to be a $\sim 3 \%$ effect for \planck \ bands. 

In the following we report on robustness tests aimed at completing this investigation. We show that our results are robust with respect to the instrumental uncertainties, that they are insensitive to the finest details of our cluster modelling, and that they are unaffected by radio source contamination. We show that restricting the analysis to the reference homogeneous subsample of NORAS/REFLEX clusters leads to measurements fully compatible with those we obtain for the whole sample.

\subsection{Beam effects}

The beam effects studied in \citet{planck2011-5.1a} are further scrutinised by directly estimating their impact on our results. To this end, the whole analysis is redone by assuming different beam FWHM. For simplicity, we systematically increase/decrease the adopted beam FWHM for all channels simultaneously by adding/removing the conservative uncertainties given in Table~\ref{tab:fwhm} from the fiducial beam FWHM values. We find that the binned SZ signal varies by at most 2\% from the value computed using the fiducial beams FWHM. 

\subsection{Modelling}

The effects of changes in the underlying X-ray based model on our results are investigated by 
repeating the full analysis as for the {\it fiducial case\/}, but by assuming the standard slope of 
the $M_{\rm 500}-Y_{\rm X}$ relation and/or the intrinsic $L_{\rm 500}$ -- $M_{\rm 500}$ relation (see Sect. \ref{clustermodel.sec}). 
For simplicity, in the following we discuss results obtained by varying only one assumption at a time. We find that 
the effect resulting by varying both assumptions is equivalent to the sum of the effects obtained by varying the two assumptions separately. 

\begin{figure}
\includegraphics[width=1.000 \columnwidth]{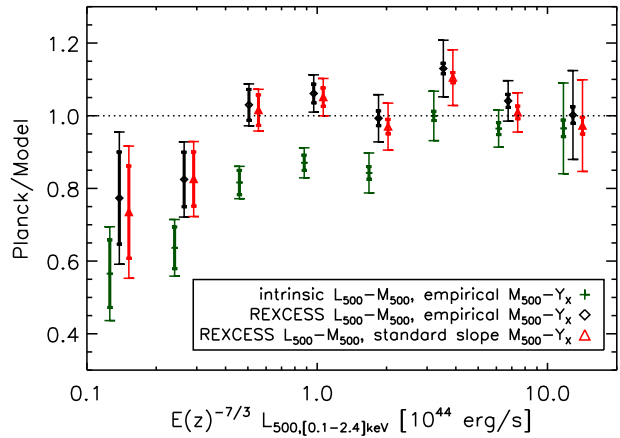}
 \caption{{\footnotesize Ratio of binned \planck \ data points to model for different model assumptions. The {\it fiducial model\/} (black diamonds) is shown together 
 with results obtained by varying the underlying $L_{\rm 500}$ -- $M_{\rm 500}$ relation from \rexcess \ to intrinsic (green plus signs), and by varying the slope of the underlying $M_{\rm 500}-Y_{\rm X}$ relation from empirical to standard (red triangles). Thick bars give the statistical errors, while the thin bars are bootstrap uncertainties.}}
 \label{cases:fig}
  \end{figure}

As expected from the weak dependence of cluster size $R_{\rm 500}$ on luminosity, the measured SZ signal is barely affected by these changes. If the standard slope case is adopted instead of the empirical one, the bin 
averaged SZ signal changes by less than a few percent at all luminosities and the same is found when the intrinsic $L_{\rm 500}$ -- $M_{\rm 500}$ relation is adopted. 
The model predictions are of course more affected by changes in the assumed scaling relations. In Fig. \ref{cases:fig} we contrast the \planck \ -to-model ratio obtained for the different cases. The figure shows that the assumption on the slope of the $M_{\rm 500}-Y_{\rm X}$ relation has a fully negligible impact. On the other hand it shows that the intrinsic $L_{\rm 500}$ -- $M_{\rm 500}$ relation is not compatible with our measurements ($> 5 \sigma$ discrepancy). This finding is not surprising given the fact that when adopting the intrinsic $L_{\rm 500}$ -- $M_{\rm 500}$ relation one assumes that selection effects of our X-ray sample are negligible. Notice that the {\it WMAP\/}-5yr data used in the similar analysis by \cite{melin2010} did not have sufficient depth to come to this conclusion. Furthermore, the agreement of our results for the \rexcess \ and intrinsic $L_{\rm 500}$ -- $M_{\rm 500}$ relations at high luminosity confirms that Malmquist bias is small for very luminous objects.  

\subsection{Intrinsic dispersion in the $L_{\rm 500}$ -- $M_{\rm 500}$ relation}

The intrinsic dispersion in the $L_{\rm 500}$ -- $M_{\rm 500}$ relation dominates the uncertainty on the clusters' size $R_{\rm 500}$ in our analysis. We investigate how this propagates into the uncertainties on the binned SZ signal by means of a Monte Carlo (MC) analysis of 100 realisations. We use the  dispersion given in Table \ref{tab:lx_m_param} and, for each realisation, we draw a random mass $\log M_{\rm 500}$ for each cluster from a Gaussian distribution with mean given by the  $L_{\rm 500}-M_{\rm 500}$ relation and standard deviation $\sigma_{\log L - \log M}/\alpha_M$. For each realisation, we extract the signal with the new values of $M_{\rm 500}$ (thus $R_{\rm 500}$). The standard deviation of the SZ signal for the 100 MC realisations in a given luminosity bin is found to be at most $\sim 3 \%$  of the signal. Hence, given the size of the total errors on the binned SZ signal (see Fig. \ref{errors:fig}) our conclusions are fully unaffected by this effect.

\subsection{Pressure profile}

Furthermore, we investigate how the uncertainties on the assumed pressure profile propagates into the uncertainties on the binned SZ signal. For simplicity, we only quantify the largest possible effect by redoing the analysis but adopting the pressure profile parameters for the cool-core and morphologically disturbed subsamples given in Table \ref{tab:pressure_param}, i.e. we assume that all clusters in the sample are cool-core or morphologically disturbed. Both of the two resulting sets of binned SZ signal deviate from the one derived assuming the universal pressure profile by approximatively $8 \%$ in the lowest luminosity bin and decrease linearly with increasing ${\rm log} L_{\rm 500}$, becoming approximatively $1 \%$ in the highest luminosity bin. Furthermore the normalisation of Eq. \ref{YM} changes by less than $ 3 \%$ if the average pressure profiles parameters of cool-core and morphologically disturbed clusters given in Table \ref{tab:pressure_param} are adopted instead of the ones for the average profile, implying that our SZ signal predictions are robust. Considering the total errors and their trend with luminosity, we conclude that our findings are fully unaffected by the exact shape of the SZ template.

\subsection{X-ray sample}

Because of the reasons detailed in Sect. \ref{X-raydata.sec}, we also repeated our analysis by considering the NORAS/REFLEX {\it control sample\/} and find results fully consistent with those derived for the full sample. The results are shown in terms of \planck \ -to-model ratio in Fig. \ref{wmap_comp:fig}. Notice that in this case the luminosity binning is chosen so as to be comparable with that in the {\it WMAP\/}-5yr analysis of \cite{melin2010}. The comparison between {\it WMAP\/}-5yr and \planck \ results is discussed in Sec. \ref{conclusions.sec} below. 

\subsection{Radio contamination}

In addition we investigated the effect of contamination by radio sources on our results. Most radio sources are expected to have a steep spectrum and hence they should not have significant fluxes at \planck \ frequencies. However, some sources will show up in \planck \ LFI and HFI channels if their radio flux is sufficiently high and/or their spectral index is near zero or positive. Extreme examples are the Virgo and Perseus clusters that host in their interior two of the brightest radio sources in the sky. In the ESZ sample \citep{planck2011-5.1a} there are also a few examples of clusters with moderate radio sources in their vicinity (1 Jy or less in NVSS) and still significant signal at LFI (and even HFI) frequencies. To check for possible contamination we combine data from SUMSS \citep[a catalog of radio sources at 0.85 GHz,][]{bock1999} and NVSS \citep[a catalog of radio sources at 1.4 GHz,][]{condon1998}.  We have looked at the positions of the clusters in our sample and searched for radio sources in a radius of 5 arcmin from the cluster centre. We find that 74 clusters have a radio source within this search radius in NVSS or SUMSS with a flux above 1 Jy. Among these, eight have fluxes larger than 10 Jy, two sources larger than 100 Jy and one is an extreme radio source with a flux larger than 1 KJy.

As a robustness test, we investigate the impact of contamination by radio sources on our results by excluding clusters hosting radio sources with fluxes larger than 1 or 5 Jy and comparing the results to those obtained for the full sample. Interestingly we find that, as expected, the individual SZ signal is on average lower than the X-ray based predictions in clusters that are likely to be highly contaminated. This is shown in Fig. \ref{planck_model:fig} where clusters associated with radio sources with fluxes larger than 5 Jy are shown by the red symbols. However, given the very low fraction of possibly contaminated clusters, we find that bin averaged signal is fully unaffected when these are excluded from the analysis.

\section{Discussion and conclusions}
\label{conclusions.sec}

As part of a series of papers on \planck\ early results on clusters of galaxies \citep{planck2011-5.1a,planck2011-5.1b,planck2011-5.2a,planck2011-5.2b,planck2011-5.2c}, we measured the SZ signal in the direction of $\sim$ 1600 objects from the MCXC \cite[Meta-Catalogue of X-ray detected Clusters of galaxies,][see Sect. \ref{X-raydata.sec}]{MCXC} in \planck \ whole sky data (see Sect. \ref{SZdata.sec}) and studied the relationship between X-ray luminosity and SZ signal strength. 

For each X-ray cluster in the sample the amplitude of the SZ signal is fitted by fixing the cluster position and size to the X-ray values and assuming a template derived from the universal pressure profile of \cite{arnaud2010}. 
The universal pressure profile was derived from high quality data from \rexcess . Recently, 
\cite{sun2010} found that the universal pressure profile also yields an excellent description of systems with lower luminosities  than those probed with \rexcess . This implies that the adopted SZ template is suitable for the entire luminosity range explored in our work.

The {\it intrinsic\/} SZ signal $D^2_{\rm A} \, Y_{\rm 500}$ is averaged in X-ray luminosity bins to maximise the statistical significance. The signal is detected at high significance over the X-ray luminosity range $10^{43} {\rm erg/s} \lesssim L_{\rm 500} E(z)^{-7/3} \lesssim 10^{45} {\rm erg/s}$ (see Fig.  \ref{y5r500:fig}). 

We find excellent agreement between observations and predictions based on X-ray data, as shown in Fig. \ref{planckvsmodel:fig}. Our results do not agree with the claim, based on a recent {\it WMAP\/}-7yr data analysis, that X-ray data over-predict the SZ signal \citep{komatsu2010}. Due to the large size and homogeneous nature of the MCXC, and the exceptional internal consistency of our cluster model, we believe that our results are very robust. Moreover, as reported in Sect. \ref{robustness.sec}, we show that our findings are insensitive to the details of our cluster modelling. Furthermore, we have shown that our results are robust against instrumental (calibration, colour correction, beam) and astrophysical (radio contamination) uncertainties.

\begin{figure}
\includegraphics[width=1.000 \columnwidth]{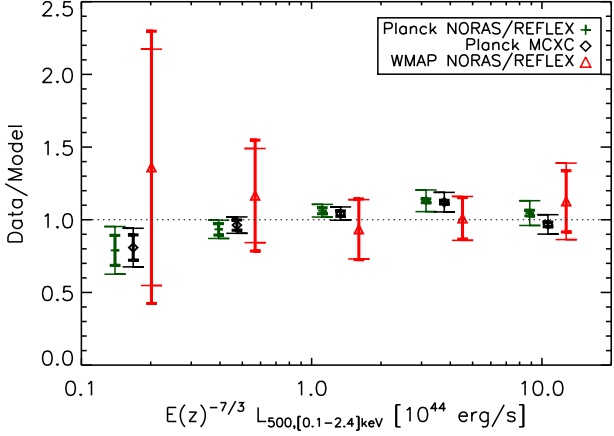}
 \caption{{\footnotesize Data-to-model ratio for \planck \ results for the full sample (black diamonds) and the NORAS/REFLEX control sample (green plus signs). The {\it WMAP\/}-5yr results for the NORAS/REFLEX by \cite{melin2010} are shown by the red triangles. Error bars are as in Fig. \ref{planckvsmodel:fig}.}}
    \label{wmap_comp:fig}
  \end{figure}

Our results confirm to a higher significance the results of the analysis by \cite{melin2010} based on {\it WMAP\/}-5yr data. This is shown in Fig. \ref{wmap_comp:fig} where the data-to-model ratio as a function of luminosity is presented. Luminosity bins are chosen so as to be comparable to those of \cite{melin2010}, and the \planck \ results are presented for the whole sample used in this work and also for the NORAS/REFLEX sample adopted in  \cite{melin2010}. In addition to the good agreement between results from the two data sets, the figure shows that in the {\it WMAP\/}-5yr study by \cite{melin2010} statistical uncertainties are dominant. As shown in Sect. \ref{binnedfluxes},  \planck \ data allows us to overcome this limitation and to investigate the intrinsic scatter in the scaling relation between intrinsic SZ signal $D^2_{\rm A} \, Y_{\rm 500}$ and X-ray luminosity $L_{\rm 500}$ (see Sect. \ref{scatter.sec}). We find a $\sim 40 \%$ intrinsic scatter in the $D^2_{\rm A} \, Y_{\rm 500}$ -- $L_{\rm 500}$ relation and show that it is linked to cluster dynamical state.

The agreement between luminosity binned X-ray predictions and \planck \ measurements is reflected in the excellent accord between predicted scaling relation and best fitting power law model to the $D^2_{\rm A} \, Y_{\rm 500}$ -- $L_{\rm 500}$ relation. The power law fit, which is performed on individual data points, is compared by \cite{planck2011-5.2b} to the {\it calibration\/} derived from a sample of galaxy clusters detected at high signal to noise in the \planck \ survey \citep[the ESZ sample, see][]{planck2011-5.1a} and with high quality X-ray data from {\it XMM-Newton\/}. As discussed in \cite{planck2011-5.2b} the slight differences between the two best fitting relations reflect the difference between the selection of the adopted samples. Indeed, the X-ray sample used in the present work is X-ray selected and therefore biased towards the cool core systems, while the sample used in \cite{planck2011-5.2b} is SZ selected.

\begin{figure}
\includegraphics[width=1.000 \columnwidth]{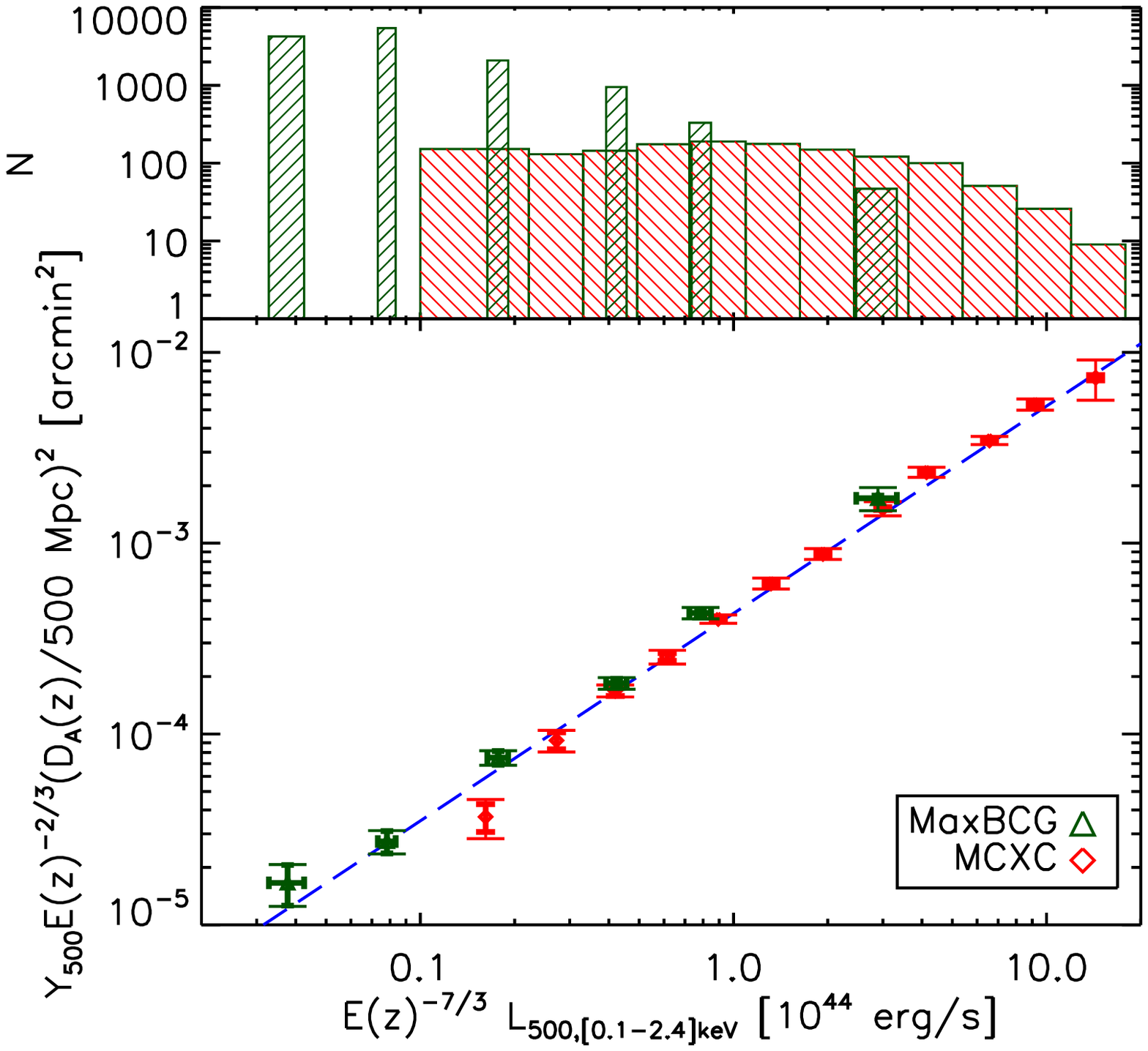}
 \caption{{{\footnotesize \it Bottom panel:\/} Comparison between our results (red diamonds, as in left-hand panel Fig. \ref{planckvsmodel:fig}) and those obtained by \cite{planck2011-5.2c} (green triangles), where MaxBCG clusters are investigated. X-ray luminosities and associated error bars for the MaxBCG clusters are based on the analysis of \cite{rykoff2008}. Vertical error bars are as in Fig. \ref{planckvsmodel:fig} and the X-ray prediction (i.e., Eq. \ref{YL}) is shown by the dashed blue line. {\it Top panel:\/} X-ray luminosity histograms of the MCXC (red) and MaxBCG (green) samples. For the MCXC the width of the bars is equal to the luminosity bin width, while for the MaxBCG we adopt the horizontal error bar shown in the bottom panel.}}
    \label{optical_comp:fig}
  \end{figure}

As mentioned in Sect. \ref{binnedfluxes}, the luminosity range where we are not able to detect the SZ signal because of the small number of low mass objects (see Fig.  \ref{y5r500:fig}), is explored in \cite{planck2011-5.2c}. In the latter analysis we use the optical catalogue of $\sim$ 14,000 MaxBCG clusters \citep{koester07} and, in a fully similar way as done in this work, extract the optical richness binned SZ signal from  \planck \ data. By combining these results with the X-ray luminosity of the MaxBCG clusters measured by \cite{rykoff2008} by stacking RASS data, in \cite{planck2011-5.2c} we derive the $D^2_{\rm A} \, Y_{\rm 500}$ -- $L_{\rm 500}$ relation for the MaxBCG sample. This result is shown together with the one derived in the present paper in Fig. \ref{optical_comp:fig}. The X-ray luminosity histograms shown in the top panel of the figure highlight the complementarity of the two analyses. The bottom panel of the figure shows agreement between the results from the two data sets and, very importantly, that observations and predictions based on X-ray data agree over a very wide range in X-ray luminosity.

We investigate the evolution of the scaling relation and find it to be consistent with standard evolution. Although redshift binned measurements and predictions agree quite well over a wide redshift range (see Fig. \ref{y500z:fig}), our constraints are weak because the inferred best fitting model is almost completely constrained by only the low redshift measurements. Given the relevance of SZ-selected samples for cosmological studies and the need for complementary X-ray observations for such studies \citep[see][and discussion therein]{planck2011-5.1b}, improved understanding of the evolution of SZ-X-ray scaling relations is clearly desired. High quality data similar to those used in \cite{planck2011-5.2b}, but for higher redshift clusters will provide tight constrains on evolution, in particular when newly SZ discovered clusters \citep[see][and references therein]{planck2011-5.1a} with high quality X-ray and optical data are included.


\begin{acknowledgements}
This research has made use of the X-Rays Clusters Database (BAX) which is operated by the Laboratoire d'Astrophysique de Tarbes-Toulouse (LATT), under contract with the Centre National d'Etudes Spatiales (CNES). We acknowledge the use of the HEALPix package \citep{2005ApJ...622..759G}. The Planck Collaboration acknowledges the support of: ESA; CNES and CNRS/INSU-IN2P3-INP (France); ASI, CNR, and INAF (Italy); NASA and DoE (USA); STFC and UKSA (UK); CSIC, MICINN and JA (Spain); Tekes, AoF and CSC (Finland); DLR and MPG (Germany); CSA (Canada); DTU Space (Denmark); SER/SSO (Switzerland); RCN (Norway); SFI (Ireland); FCT/MCTES (Portugal); and DEISA (EU). A description of the Planck Collaboration and a list of its members, indicating which technical or scientific activities they have been involved in, can be found at  \url{http://www.rssd.esa.int/Planck}. 
\end{acknowledgements}

 
\bibliographystyle{aa} 
\bibliography{Planck2011-5.2a_bib.bib,Planck_bib.bib} 
\end{document}